\documentclass[12pt]{article}
\usepackage{epsfig}
\usepackage{amsmath}
\usepackage{hhline}
\usepackage{amssymb}
\usepackage{times}

\newlength{\dinwidth}
\newlength{\dinmargin}
\begin{document}  
%
%
%
%
%
%
\newcommand{\pom}{{I\!\!P}}
\newcommand{\slowpi}{\pi_{\mathit{slow}}}
\newcommand{\fiidiii}{F_2^{D(3)}}
\newcommand{\fiidiiiarg}{\fiidiii\,(\beta,\,Q^2,\,x)}
\newcommand{\n}{1.19\pm 0.06 (stat.) \pm0.07 (syst.)}
\newcommand{\nz}{1.30\pm 0.08 (stat.)^{+0.08}_{-0.14} (syst.)}
\newcommand{\fiidiiiful}{F_2^{D(4)}\,(\beta,\,Q^2,\,x,\,t)}
\newcommand{\fiipom}{\tilde F_2^D}
\newcommand{\ALPHA}{1.10\pm0.03 (stat.) \pm0.04 (syst.)}
\newcommand{\ALPHAZ}{1.15\pm0.04 (stat.)^{+0.04}_{-0.07} (syst.)}
\newcommand{\fiipomarg}{\fiipom\,(\beta,\,Q^2)}
\newcommand{\pomflux}{f_{\pom / p}}
\newcommand{\nxpom}{1.19\pm 0.06 (stat.) \pm0.07 (syst.)}
\newcommand {\gapprox}
   {\raisebox{-0.7ex}{$\stackrel {\textstyle>}{\sim}$}}
\newcommand {\lapprox}
   {\raisebox{-0.7ex}{$\stackrel {\textstyle<}{\sim}$}}
\def\gsim{\,\lower.25ex\hbox{$\scriptstyle\sim$}\kern-1.30ex%
\raise 0.55ex\hbox{$\scriptstyle >$}\,}
\def\lsim{\,\lower.25ex\hbox{$\scriptstyle\sim$}\kern-1.30ex%
\raise 0.55ex\hbox{$\scriptstyle <$}\,}
\newcommand{\pomfluxarg}{f_{\pom / p}\,(x_\pom)}
\newcommand{\dsf}{\mbox{$F_2^{D(3)}$}}
\newcommand{\dsfva}{\mbox{$F_2^{D(3)}(\beta,Q^2,x_{I\!\!P})$}}
\newcommand{\dsfvb}{\mbox{$F_2^{D(3)}(\beta,Q^2,x)$}}
\newcommand{\dsfpom}{$F_2^{I\!\!P}$}
\newcommand{\gap}{\stackrel{>}{\sim}}
\newcommand{\lap}{\stackrel{<}{\sim}}
\newcommand{\fem}{$F_2^{em}$}
\newcommand{\tsnmp}{$\tilde{\sigma}_{NC}(e^{\mp})$}
\newcommand{\tsnm}{$\tilde{\sigma}_{NC}(e^-)$}
\newcommand{\tsnp}{$\tilde{\sigma}_{NC}(e^+)$}
\newcommand{\st}{$\star$}
\newcommand{\sst}{$\star \star$}
\newcommand{\ssst}{$\star \star \star$}
\newcommand{\sssst}{$\star \star \star \star$}
\newcommand{\tw}{\theta_W}
\newcommand{\sw}{\sin{\theta_W}}
\newcommand{\cw}{\cos{\theta_W}}
\newcommand{\sww}{\sin^2{\theta_W}}
\newcommand{\cww}{\cos^2{\theta_W}}
\newcommand{\trm}{m_{\perp}}
\newcommand{\trp}{p_{\perp}}
\newcommand{\trmm}{m_{\perp}^2}
\newcommand{\trpp}{p_{\perp}^2}
\newcommand{\alp}{\alpha_s}
\newcommand{\alps}{\alpha_s}
\newcommand{\sqrts}{$\sqrt{s}$}
\newcommand{\LO}{$O(\alpha_s^0)$}
\newcommand{\Oa}{$O(\alpha_s)$}
\newcommand{\Oaa}{$O(\alpha_s^2)$}
\newcommand{\PT}{p_{\perp}}
\newcommand{\JPSI}{J/\psi}
\newcommand{\sh}{\hat{s}}
\newcommand{\uh}{\hat{u}}
\newcommand{\MP}{m_{J/\psi}}
\newcommand{\PO}{I\!\!P}
\newcommand{\xbj}{x}
\newcommand{\xpom}{x_{\PO}}
\newcommand{\ttbs}{\char'134}
\newcommand{\xpomlo}{3\times10^{-4}}  
\newcommand{\xpomup}{0.05}  
\newcommand{\dgr}{^\circ}
\newcommand{\pbarnt}{\,\mbox{{\rm pb$^{-1}$}}}
\newcommand{\gev}{\,\mbox{GeV}}
\newcommand{\WBoson}{\mbox{$W$}}
\newcommand{\fbarn}{\,\mbox{{\rm fb}}}
\newcommand{\fbarnt}{\,\mbox{{\rm fb$^{-1}$}}}
\newcommand{\qsq}{\ensuremath{Q^2} }
\newcommand{\gevsq}{\ensuremath{\mathrm{GeV}^2} }
\newcommand{\et}{\ensuremath{E_t^*} }
\newcommand{\rap}{\ensuremath{\eta^*} }
\newcommand{\gp}{\ensuremath{\gamma^*}p }
\newcommand{\dsiget}{\ensuremath{{\rm d}\sigma_{ep}/{\rm d}E_t^*} }
\newcommand{\dsigrap}{\ensuremath{{\rm d}\sigma_{ep}/{\rm d}\eta^*} }
\newcommand{\etsq}{\ensuremath{E_t^{*2}} }
\newcommand{\etbr}{\ensuremath{\overline{E_t}} }
\newcommand{\etbarsq}{\ensuremath{\overline{E_t}^2} }
\newcommand{\xgjets}{\ensuremath{x^{jets}_{\gamma}} }
\newcommand{\xgrec}{\ensuremath{x^{rec}_{\gamma}} }
\newcommand{\xg}{\ensuremath{x_{\gamma}} }
\newcommand{\xgamma}{\ensuremath{x_{\gamma}} }
\newcommand{\triple}{\ensuremath{{\rm d}^3\sigma_{ep}/{\rm d}\qsq{\rm
      d}\etbarsq{\rm d}\xgjets} }
\newcommand{\tripart}{\ensuremath{{\rm d}^3\sigma_{ep}/{\rm d}\qsq{\rm
      d}P_t^2{\rm d}x_{\gamma}}}
\def\Journal#1#2#3#4{{#1} {\bf #2}, #3 (#4)}
\def\NCA{\em Nuovo Cimento}
\def\NIM{\em Nucl. Instrum. Methods}
\def\NIMA{{\em Nucl. Instrum. Methods} A}
\def\NPB{{\em Nucl. Phys.} B}
\def\PLB{{\em Phys. Lett.}  B}
\def\PRL{\em Phys. Rev. Lett.}
\def\PRD{{\em Phys. Rev.} D}
\def\ZPC{{\em Z. Phys.} C}
%
%
%
%
%
%
\begin{titlepage}
\begin{flushleft}
DESY-98-205 \hfill ISSN 0418-9833\\
December 1998
\end{flushleft}

\vspace{2cm}

\begin{center}
\begin{Large}

  {\bf Measurement of Dijet Cross-Sections
    at Low ${\rm{\bf Q^2}}$ \\
    and the\\
    Extraction of an Effective Parton Density \\
    for the Virtual Photon}

\vspace{1cm}

H1 Collaboration

\end{Large}
\end{center}

\vspace{1cm}

\begin{abstract}
\noindent
 The triple-differential dijet cross-section, $\triple$,
 is measured with the H1 detector at HERA 
 as a function of the photon virtuality $Q^2$, the
 fraction of the photon's momentum carried by the parton entering the
hard scattering, $\xgjets$, and the
 square of the mean transverse energy, $\etbarsq$, of the two highest $E_t$ jets. 
Jets are
 found using a longitudinal boost-invariant $k_T$ clustering
 algorithm in the $\gamma^* p$ center of mass frame.
 The measurements cover the ranges  $1.6 < Q^2 < 80$\, GeV$^2$ 
 in virtuality and $0.1 < y < 0.7$ in inelasticity $y$.
 The results are well described by
 leading order QCD models which include
 the effects of a resolved component to the virtual photon. Models which
 treat the photon as point-like fail to describe the data.
 An effective leading order parton density for the virtual photon
 is extracted as a function of the photon virtuality, the 
 probing scale and the parton momentum fraction. The
 $x_{\gamma}$ and probing scale dependences of the parton density 
 show characteristic features of
 photon structure, and a suppression of this structure with increasing $Q^2$ is seen. 
\end{abstract}

\vspace{1.5cm}
\begin{center}
Submitted to {\it Eur. Phys. J. C.}
\end{center}
\end{titlepage}
%
%
%
%
%
%
\noindent
\begin{flushleft}
 C.~Adloff$^{34}$,                
 V.~Andreev$^{25}$,               
 B.~Andrieu$^{28}$,               
 V.~Arkadov$^{35}$,               
 A.~Astvatsatourov$^{35}$,        
 I.~Ayyaz$^{29}$,                 
 A.~Babaev$^{24}$,                
 J.~B\"ahr$^{35}$,                
 P.~Baranov$^{25}$,               
 E.~Barrelet$^{29}$
% ...


\end{flushleft}
\begin{flushleft}
\noindent
% ...
$ ^2$ III. Physikalisches Institut der RWTH, Aachen, Germany$^a$ \\
% ...
$ ^{24}$ Institute for Theoretical and Experimental Physics,
          Moscow, Russia \\
 $ ^{25}$ Lebedev Physical Institute, Moscow, Russia$^{f,k}$ \\
% ...
$ ^{28}$ LPNHE, \'{E}cole Polytechnique, IN2P3-CNRS, Palaiseau, France \\
 $ ^{29}$ LPNHE, Universit\'{e}s Paris VI and VII, IN2P3-CNRS,
          Paris, France \\
% ...
$ ^{34}$ Fachbereich Physik, Bergische Universit\"at Gesamthochschule
          Wuppertal, Wuppertal, Germany$^a$ \\
 $ ^{35}$ DESY, Institut f\"ur Hochenergiephysik, Zeuthen, Germany$^a$ \\
% ...
$ ^a$ Supported by the Bundesministerium f\"ur Bildung, Wissenschaft,
        Forschung und Technologie, FRG,
        under contract numbers 7AC17P, 7AC47P, 7DO55P, 7HH17I, 7HH27P,
        7HD17P, 7HD27P, 7KI17I, 6MP17I and 7WT87P \\
% ...
$ ^f$ Supported by the Deutsche Forschungsgemeinschaft \\
% ...
$ ^k$ Supported by Russian Foundation for Basic Research 
       grant no. 96-02-00019 

\end{flushleft}
%
%
%
%
%
%
%
%
%
%
\newpage
\section{Introduction and Motivation}
It is
well-established that the real photon has a partonic structure 
both through measurements in two-photon
collisions at electron-positron colliders~\cite{ggref1} and through measurements of
the photoproduction of jets at HERA~\cite{herajets,Rick,h1eflow}. These data have been used to 
determine universal parton densities for real photons. 
The dynamical evolution of the partonic structure of the photon as it becomes
virtual is described in a QCD framework both for
deep-inelastic positron-proton (ep) collisions and for high transverse
momentum ($P_t$) processes in two-photon and 
photoproduction reactions~\cite{vpth1}.
Experimental data with target photons of sizeable virtuality in two-photon collisions are, however,
sparse~\cite{ggmax}. The ep 
collision data at HERA on the other hand
are available over
a wide range of photon virtualities, $Q^2$, from photoproduction to 
high $Q^2$ deep-inelastic scattering (DIS) and are 
sensitive to any photon structure~\cite{vpth2,kkp,vpth3}.

The production of high transverse energy ($E_t$) jets in ep
collisions is dominated by processes
in which a single space-like photon carries the momentum transfer
from the incident positron. When this photon is quasi-real, i.e. in 
photoproduction processes, two types of interaction can be distinguished in
leading order:
direct processes in which the photon couples as a point-like object to
a parton out of the proton  and resolved processes in which it
develops a partonic structure prior to the collision. In the latter case, 
a parton out of the photon, carrying only a fraction 
$\xg$ of the photon momentum, enters the hard 
scattering process leading to the production of jets. 
Examples of these two types of process are shown
in figure 1.
The photoproduction jet cross-section is 
therefore sensitive to the density of partons in the photon and
the latter can be measured. 
A natural choice of the scale, $P_t$, at which this 
photon structure is probed is given
by the $E_t$ of the jets with respect to the $\gamma p$ axis 
in the $\gamma p$ centre of mass system (cms).

The diagrams in figure 1 are equally applicable to processes
where the exchanged photon is highly virtual. In such deep-inelastic
scattering processes,
the production of high $E_t$ jets is usually
dominated by direct processes.
However, 
as long as the photon
is probed with sufficiently high resolution, i.e. if $P_t^2 \gg Q^2$ (with 
$P_t$ again defined by the $E_t$ of the jets in the $\gamma^* p$ frame), 
it may still be possible to have interactions in which the cross-section
factorises and the photon structure 
is resolved~\cite{vpth2,vpth3,vpth4,vpth5,vpth6}.  
It is then a prediction of perturbative QCD that
the parton densities of virtual photons become suppressed as
$Q^2$ increases at fixed $P_t^2$.  The partonic 
structure in the photon becomes simpler, 
until only the direct coupling 
to a ${\rm q} \overline{{\rm q}}$ pair
remains~\cite{vpth2,vpth3,vpth4,vpth5,vpth6}. 
The concept of virtual photon structure thus provides a unified 
description of high $E_t$ jet production over the whole range of $Q^2$
corresponding to a smooth transition between photoproduction and DIS processes.
It is implicit in this concept that the structure is universal. Future
comparisons with, for example, $\gamma^* \gamma$ data may establish whether
this is the case.

In a previous publication~\cite{loq2jet}, it was shown that the single inclusive 
jet cross-section in low $\qsq$ deep-inelastic scattering 
can indeed be described by models which include a 
contribution from resolved virtual photons which is suppressed 
with increasing $Q^2$. 

The dijet cross-section has also been measured
in low $Q^2$ deep-inelastic scattering processes and has been found to be best
described with predictions in which the effects of virtual photon 
structure are included~\cite{grind}.

With dijet events, it becomes possible to reconstruct the variable
$\xg$ and hence extract an effective parton density for the photons. 
Such a measurement has recently been made using photoproduction 
events~\cite{Rick}.

\begin{sloppypar}
In this paper we extend the latter studies
and investigate the evolution
of the effective parton density with $Q^2$ as well as the $P_t^2$ of the
partons. 
We begin by measuring the 
triple-differential jet cross-section, 
$\triple,$ where $\overline{E_t}$ is 
the mean of the transverse energies of the two highest $E_t$ 
jets measured in the $\gamma^* p$ centre of mass frame, and $\xgjets$ is 
the value of $\xg$ as estimated from the jets.
Jets were found using
the inclusive $k_T$ algorithm~\cite{inckt}. 
The measured cross-section is
compared to simulations with  LO matrix elements
which 
include models for the virtual photon structure. 
\end{sloppypar}

The cross-section is then used 
to extract a leading order effective 
parton density as a function of $x_{\gamma}$, $Q^2$
and $P_t^2$. The observed shape, scaling behaviour and virtuality dependences
of this effective parton density are compared with various parameterisations
based on predictions from perturbative QCD.
\section{The H1 Detector}
The H1 detector is described in detail in~\cite{h1det}. In this analysis, we
make particular use of the SPACAL~\cite{spacal} calorimeter for detection and identification of the 
scattered positron. The hadronic energy flow is measured with the Liquid
Argon (LAr)~\cite{Lar} and SPACAL calorimeters. 
The central and forward tracking detectors are used to 
reconstruct the event vertex and to supplement the measurement of hadronic energy
flow made with the LAr and SPACAL.

We use a coordinate system in which the nominal interaction point is at
the origin and the incident proton beam defines the $+z$ direction.
The polar angle $\theta$ is defined with respect to the proton
direction.
 
The central tracking system consists of two concentric cylindrical drift
chambers, coaxial with the beam-line and centered about the nominal
interaction vertex. Its polar angle coverage, 
$15^\circ < \theta <  165^\circ$, is complemented by that of the
forward tracker, $7^\circ < \theta < 25^\circ$. The central tracker is
interleaved with drift chambers providing measurement of the $z$ coordinates
of the tracks. The tracking detectors are immersed in a 1.15 T 
magnetic field generated by a superconducting solenoid which surrounds the
LAr.
The LAr is a finely grained calorimeter
covering the range in polar angle
$4^\circ < \theta <  154^\circ$  
with full azimuthal
acceptance. It consists of an electromagnetic section with 
lead absorbers, 20--30 radiation lengths in depth,
and a hadronic section with steel absorbers. The
total depth of the calorimeter varies between
4.5 and 8 hadronic interaction lengths. 
The energy resolution is  $\sigma(E) / E \approx
0.12 / \sqrt{E}\oplus 1\%$ for positrons and $\sigma(E) / E \approx 0.5 /
\sqrt E\oplus 2\% $
for hadrons ($E$ in GeV), as measured in test beams~\cite{tbeam}. 
The absolute energy scale is known to a 
precision of  3\% for positrons and 4\% for hadrons.

\begin{sloppypar}
The SPACAL is a lead/scintillating-fibre calorimeter which covers the 
angular region \mbox{$153^\circ < \theta < 177.8^\circ$}. It contains an electromagnetic
and a hadronic section. The former has an energy resolution of
$7.5\% / \sqrt{E}\oplus 2.5\%$. The energy resolution for hadrons is~$\sim 30\%
/ \sqrt{E}$. The SPACAL provides the main trigger for the 
events (see section \ref{evsel}) in this analysis. 
The timing resolution of better than 1 ns in both sections of the SPACAL 
is exploited in the trigger to reduce proton beam induced background. 
The energy scale uncertainty of the electromagnetic section is 2\% and 
that of the hadronic section is 7\%.
The backward drift chamber (BDC) system in front of the SPACAL spanning the
region $151^\circ< \theta < 177.5^\circ$
provides track segment information to improve positron identification in the 
SPACAL. In conjunction with the event vertex determination from the
central and forward track detectors, it gives a precision measurement of the
positron scattering angle of 1 mrad.
\end{sloppypar}

The luminosity determination is based on measurement of the ${\rm ep}\rightarrow
{\rm ep}\gamma$ Bethe-Heitler process.
The positron and photon are detected in the electron tagger
located at $z=-33.4$ m and photon tagger at $z=-103$ m, respectively. Both consist
of crystal Cherenkov calorimeters with a resolution of $\sigma(E)/E \approx
0.1/\sqrt{E}$. The integrated luminosity was measured to a precision of better
than 2\%.

\section{Theoretical Models and Simulations}
\subsection{Monte Carlo Models}
The analysis uses simulated events both to correct the measured
cross-sections for detector effects and in order to compare the data
with the predictions of various theoretical models. 
The various combinations of Monte Carlo simulations and parton densities
used in this analysis are summarised in table~\ref{Tab1}.

The HERWIG~\cite{herwig} and RAPGAP~\cite{rapgap}
Monte Carlo models are both able to simulate the direct and resolved production of
dijets by virtual photons. 
In both models the hard scattering process is simulated in leading
order (LO), regulated with a minimum $P_t$ cut-off, $P_t^{min}$,
and supplemented by initial and final-state parton showers. 
In the simulation of resolved processes, the
equivalent photon approximation is used for the flux of transversely
polarised  photons and on-shell $2\rightarrow 2$ matrix elements are taken. The
longitudinal flux is not included. Exact ${\rm eq} \rightarrow {\rm eqg}$
and ${\rm eg} \rightarrow {\rm eq}\overline{{\rm q}}$ matrix elements are used for the
direct photon component. 

HERWIG has parton showering based on colour
coherence and uses the 
cluster model for hadronisation~\cite{herps}. Some tuning of the parton showering
is possible and we use two different settings for the scale\footnote{Specified by the parameter
QSPAC.} at which the parton showering is terminated. In addition to the hard scattering process, 
HERWIG can also model the additional soft underlying activity in the
event (SUE) which is necessary to describe the observed energy flow in and around 
the jets~\cite{h1eflow}.
Such soft particles (uncorrelated with the hard scattering process) may be
produced via soft remnant-remnant interactions. 
The probability that a resolved event contains soft underlying activity was
adjusted in the simulation. Event samples with 0\% and 100\% SUE were mixed to
simulate different probabilities. No SUE was introduced into the direct sample.

RAPGAP uses a  leading-log parton shower approach and the LUND string model~\cite{jetset}
for hadronisation. It contains no mechanism for simulating
additional soft underlying activity in the events.

The simulations have interfaces to a variety of parameterisations of photon 
and proton parton density functions (PDF). 
The factorisation scales of the proton and photon parton densities were both
set equal to the $P_t$ of the scattered partons with respect to the $\gamma
p$ axis in the $\gamma p$ cms.
GRV parton densities~\cite{GRVp} were used for
the proton. When correcting the data for detector effects, 
higher order (HO)
versions of the parton densities and the 2-loop expression for $\alpha_S$ were used. As has 
also been noted by the ZEUS collaboration~\cite{z1.7}, with
this configuration it is necessary to re-scale
the HERWIG predictions by a factor of 1.7 in order to describe the data. 

Several models for the virtual photon parton densities were
considered. 
The Drees-Godbole model (DG)~\cite{vpth5, vpth6} starts with real photon parton densities~\cite{GRVgam}
and suppresses them by a factor $\mathcal{L}$ which depends on $Q^2$, $P_t^2$ 
and a free parameter, $\omega$, which controls the onset of the suppression:
\begin{equation}
\mathcal{L}(Q^2,P_t^2,\omega^2) = \frac{{\rm
    ln}\displaystyle\frac{P_t^2+\omega^2}{Q^2+\omega^2}}    {{\rm ln}\displaystyle\frac{P_t^2+\omega^2}{\omega^2}}
\end{equation}
Quark densities in the real photon are suppressed by $\mathcal{L}$ and the
gluon densities by $\mathcal{L}^2$. 
This ansatz, based on the analysis in \cite{vpth6}, is designed to interpolate
smoothly between the leading-logarithmic part of the real photon parton
densities, $\sim {\rm ln}(P_t^2/\Lambda_{QCD}^2)$, and the asymptotic domain,
$P_t^2 \gg Q^2 \gg \Lambda_{QCD}^2$, where the photon density functions are
predicted by perturbative QCD to behave as $\sim
{\rm ln }(P_t^2/Q^2)$. In this model, the shape of the $x_{\gamma}$
distribution evolves with $Q^2$ because of 
the different suppression of the quark and gluon
densities. 

In the models of Schuler and Sj{\"o}strand (SAS)~\cite{SAS}, the virtual photon
parton densities are decomposed into a non-perturbative component modelled by vector meson
dominance (VMD) and a perturbative
anomalous component. As $Q^2$ increases, the VMD component is rapidly suppressed, 
$\sim [m_V^2/(m_V^2 + Q^2)]^2$, whereas the anomalous part has a
slower logarithmic suppression and is again designed to approach the exact QCD predictions in
the $P_t^2 \gg Q^2 \gg \Lambda_{QCD}^2$ region. There are four models, SAS-1M,
SAS-2M, SAS-1D and SAS-2D
which differ in their choice of factorisation scheme (DIS~(D) or $\overline{\rm MS}$~(M)) and the scale at which
the evolution is started (0.6 or 2.0 GeV indicated by the 1 or 2 in the name, respectively).

The events used for the correction of detector effects were processed through
a full simulation of the H1 detector. The HERWIG event sample which we use as
our main model for the corrections, contains approximately 3 times as many events
as the selected data sample. The statistics of the RAPGAP sample, which we
use in the estimation of systematic errors, are
comparable to that of the data. 

\begin{table}
\begin{tabular}{|l|c|c|c|c|c|c|}
  \hline 
  Model name
  & $\alpha_s$ & Proton PDF & $\gamma^*$ PDF & ${\rm P_t^{min}}$ &
  SUE & QSPAC \ \\
  &            &    &  & (GeV) & & (GeV)  \ \\

  \hline\hline 
  HERWIG(HO)/DG &  
  2-loop & GRV-HO  & GRV-HO*DG & 3 & 10\% 
  & 1.0 \ \\
  \hfill ($\times 1.7$) & & & ($\omega = 0.2\, {\rm GeV}$) & & & \ \\
  RAPGAP(HO)/SAS-2D & 1-loop & GRV-HO & SAS-2D & 3 & - & - \  \\
  \hline
  HERWIG(LO)/DG &  1-loop & GRV-LO & GRV-LO*DG &
  2 & 5\% & 2.0 \ \\
   & & & ($\omega = 0.2\, {\rm GeV}$) & & & \ \\
  RAPGAP(LO)/DG & 1-loop & GRV-LO & GRV-LO*DG 
  & 2 & - & -
  \ \\
   & & & ($\omega = 0.2\, {\rm GeV}$) & & & \ \\
  RAPGAP(LO)/SAS-2D & 1-loop & GRV-LO & SAS-2D & 2 & - & -
  \ \\
  RAPGAP(LO)/SAS-1D & 1-loop & GRV-LO & SAS-1D & 2 & - & -
  \ \\
  \hline
\end{tabular}


\caption{Description of models. The first two models are used to correct for
  detector effects. Note that the predictions from the first model were scaled by a factor
  of 1.7. The remainder are used in the unfolding of the 
  diparton cross-section from the corrected jet cross-section
  and for comparison with the
  corrected cross-section and effective parton density.}
\label{Tab1} 
\end{table}
\section{Event Selection}
\label{evsel}
The analysis is based on positron-proton collision data 
collected by the H1 detector at HERA in 1996 and corresponds 
to an integrated luminosity of $\sim 6~{\rm pb}^{-1}$. 
During this period, 820 GeV protons collided with 27.5 GeV 
positrons. The events were triggered by an energy 
deposition exceeding a threshold energy in the  electromagnetic section of the
SPACAL of $\gsim 2.5~{\rm GeV}$ at large radii or $5.7~{\rm GeV}$ at large radii,
provided this was accompanied by at least one track
in the central tracking device with transverse momentum $P_t\gsim 0.8$ GeV and a well-defined
interaction vertex. The efficiency of this trigger is typically
$\sim 90\%$ and has been measured over the relevant kinematic range to a
 precision of 5\%.

Additional criteria were then applied to reduce background events and to ensure that
the events were well-measured. The scattered positron was identified as
follows. An energy cluster in the SPACAL was required
to have a radius of less than $3.5~{\rm cm}$, consistent with being produced by a
positron. In order that it be well-contained, the cluster was required 
to be at a radius greater than $8~{\rm cm}$, to have less than $10\%$ of its energy in the
innermost cells of the calorimeter and to have $< 0.5~{\rm GeV}$ of energy
deposited in the cells of the hadronic section immediately
behind the cluster. Finally, a track in the BDC was required such that 
the radii of the SPACAL cluster and BDC track differed by no more than $3.0~{\rm cm}$.

The reconstructed event vertex was required to lie
within 30 cm of its nominal location in $z$. In order to further reduce
photoproduction background, we required that 
$\sum_{i}(E^{i}-P_z^{i})$, which is expected to approximately  
equal twice the electron beam energy, $E_e$, for DIS events,
lies in
the range 40--65 GeV. The
sum is taken over all calorimeter clusters supplemented by tracking
information. This procedure corrects
for energy loss in the passive material in front of the calorimeter.
Each track was allowed to contribute a maximum of 0.35 GeV to
avoid double counting with the calorimetric energy. 

The inelasticity, calculated from the energy, $E^{\prime}_e$,
 and polar angle, $\theta^{\prime}$,
of the scattered positron:
\begin{equation}
y_e = 1-E^{\prime}_e/E_e\sin^2(\theta^{\prime}/2),
\end{equation}
was restricted to the range $0.1 < y_e < 0.7$. The upper limit
corresponds to a minimum scattered positron energy of $\sim 8$ GeV. Small
$y_e$ values, where the measurement is limited by the energy resolution of SPACAL,
were excluded.
The positron energy and scattering angle  were also used to calculate the virtuality, $Q^2$, of
the photon:
\begin{equation}
Q^2 = 4E^{\prime}_eE_e\cos^2(\theta^{\prime}/2).
\end{equation}

The inclusive $k_t$ clustering algorithm~\cite{inckt} was applied to find jets
in the $\gamma^{*}p$ cms frame with the boost
 defined by the scattered positron's energy and scattering angle. Tracking information was used
to improve the reconstruction of the $E_t$ of the jets as described
above. The calorimeter energy clusters 
were treated as massless objects and 
the tracks
were assigned the pion mass.
The clustering of these final state
objects into jets uses as the distance measures:
\begin{align}
d_{ij}&=\min(P_t^{i\;2}, P_t^{j\;2})\Delta R\\
{\rm where}\qquad\Delta R&= \sqrt{(\eta^i-\eta^j)^2+(\phi^i-\phi^j)^2}\\
{\rm and} \qquad d_{i}&=P_t^{i\;2}
\end{align}
between objects $i$ and $j$ and between the $i$'th object
and the beam respectively. The $P_t^i$, $\eta^i$
and $\phi^i$ are the transverse momenta, pseudorapidities given by $\eta^i =
-\ln\tan(\theta^i/2)$ and azimuthal
angles of the objects, respectively. At each iteration of the algorithm, the smallest
distance measure is determined. If this is one of the $d_{ij}$, then 
the objects $i$ and $j$ are combined into a massless object using the $P_t$ weighting scheme:
\begin{align}
P_t = P_t^i + P_t^j\\
\eta = \frac{\eta^i P_t^i + \eta^j P_t^j}{P_t}\\
\phi = \frac{\phi^i P_t^i + \phi^j P_t^j}{P_t}
\end{align}
Whenever a $d_i$ is the smallest distance measure, the
$i$'th object defines a completed jet and is
excluded from further iterations. The iterations terminate when all objects
have been assigned to jets. The reconstructed jets are massless.

Events were required to contain at least two jets found by the above
algorithm. 
Only the two highest $E_t$
jets are used in the analysis. 
Events were accepted if the two highest $E_t$
jets in the event (called in the following jet 1 and jet 2) satisfied the following criteria:
\begin{eqnarray}
|\eta^{jet\;1} -\eta^{jet\;2}| < 1.0 \\
-3.0 < \overline{\eta} < -0.5 \\
\overline{E}_t^2 > 30 \;{\rm GeV}^2 \\
\frac{|E_t^{jet\;1}-E_t^{jet\;2}|}{(E_t^{jet\;1}+E_t^{jet\;2})} < 0.25
\end{eqnarray}
where $\overline{\eta}$ and $\overline{E}_t$ are the mean pseudorapidity and
mean transverse energy of the two highest $E_t$ jets. $E_t$ and $\eta$ are 
always given with reference to the $\gamma^{*}p$ cms with the proton
direction defining the positive $z$-axis. 
The same cuts, applied to the jets after correction for detector effects, 
serve to define the cross-section. The cuts are the same as those used in
reference~\cite{Rick}.
The first two  cuts help to ensure that the jets are confined to a region of the detector
where they are well-measured and therefore that $x_{\gamma}$ can be well-determined. 
The restrictions on the difference in $\eta$ and $E_t$ of the jets
reduce the probability of misidentifying a part of the photon or proton remnant as one
of the high $E_t$ jets.
The constraints on $E_t$ are such that neither jet 
has $E_t < 4~{\rm GeV}$ and the sum of the jet $E_t$'s is always
$\gapprox 11~{\rm GeV}$. 
In 10\% of the selected events there is
a third jet with $E_t > 4~{\rm GeV}$. 
Although we do not do so here, the asymmetry in the jet 
selection makes possible comparison with
NLO QCD calculations which become unstable for symmetric jet cuts~\cite{frixrid}
if both jets are near their common lower limit. 

After application of these selection criteria we obtained a sample of 
approximately 12,000 dijet events with $\etbarsq > 30 \;{\rm GeV}^2$ 
spanning the $Q^2$ range $1.6 < Q^2 < 80\; {\rm GeV}^2$. Diffractive
events were not explicitly excluded.
The residual
background from photoproduction is negligible.

\section{Measurement of the Triple Differential Dijet Cross-section}
We first study the dependence of the dijet cross-section
on the variables $x_{\gamma}^{jets}$, $\overline{E}_T^2$ and $Q^2$. 
The jet-based variable, $x_{\gamma}^{jets}$, is related to the true 
$x_{\gamma}$ of the events:
\begin{equation}
x_{\gamma}^{jets} = 
\frac{\displaystyle{\sum_{jets\;1,\;2}} (E^{jet}-p_{z}^{jet})} {\displaystyle{\sum_{h}}(E^{h}-p_{z}^{h})}.
\label{eq:xrec}
\end{equation}
Energies and momenta are measured in the $\gamma^* p$ cms and with respect to
the $\gamma^* p$ axis. The sum in the denominator
is over all final state particles (except for the scattered positron).
As follows from the conservation of energy and longitudinal momentum,
$\xgjets$ is equal to the true $x_{\gamma}$ for leading order dijet production.
 
In each event, an 
estimate of $x_{\gamma}^{jets}$
was made using the reconstructed
energies and longitudinal momenta of the two highest $E_t$ jets in the
$\gamma^* p$ cms.  This quantity is referred to as $x_{\gamma}^{rec}$
below. The sum in the denominator was taken over all 
reconstructed objects in the event (calorimeter clusters supported 
by tracking information) except for those associated with the scattered positron.

\subsection{Correction of the Data for Detector Effects} 
The measured dijet cross-sections were corrected for detector acceptance
and resolution effects in the kinematic domain specified in the previous
section.
An iterative two-dimensional Bayesian unfolding
technique~\cite{dagostini} was applied to distributions of
$x_{\gamma}^{rec}$ and $\overline{E}_T^{2\;rec}$ in separate ranges of reconstructed $Q^2$.
The correlations between the measured variables
$\{x_{\gamma}^{rec},\overline{E}_{T}^{2\;rec}\}$ and the corrected variables
$\{x_{\gamma}^{jets},\overline{E}_T^2\}$ were obtained using events generated by
the 
HERWIG(HO)/DG model (see Table 1). 
The generated events were  
subjected to a detailed simulation of
the H1 detector.  
The $x_{\gamma}^{jets}$ resolution is approximately $12\%$, independent of
the $x_{\gamma}^{jets}$, $\etbarsq$ and $Q^2$ values. 
The resolution in
$\etbarsq$ is $\sim\!14\%$ in the highest $\etbarsq$ range
reducing in precision to $\lsim\!40\%$ in the lowest.
The resolution in $Q^2$ is $8\%$ at low $Q^2$
and $2\%$ at high $Q^2$. 
A bin-by-bin correction was then applied for the
$Q^2$ dependence. The measurement error on $Q^2$ is much smaller than the bin
size and migrations are negligible. 
After unfolding, the correlated error between bins in the unfolded distributions was $< 40\%$.
The systematic errors are described below. Here as elsewhere in this analysis, the various systematic errors are
added in quadrature and unless otherwise stated, the
numbers represent average values for the errors.
\begin{itemize}
\item {\it Model Dependence.} The largest sources of systematic error arise from model dependences, 
in particular from the choice of parton showering and 
hadronisation models.
These were estimated by comparing the results of unfolding using HERWIG with
those from unfolding 
with the RAPGAP(HO)/SAS-2D simulation and then by assigning the full difference as the error symmetrically
to the measurement.
The estimate of $\sim \pm 20\%$ includes the small contribution 
from the choice of input parton densities.
\item {\it Stability of Unfolding.} By varying the number of iterations used, we estimated that unfolding
instabilities result in a 5\% systematic uncertainty in the unfolded
cross-sections.
\item {\it Absolute Energy Scales.} The uncertainties in the LAr and SPACAL calorimeter
hadronic energy scales lead to 12\% and 1\% systematic errors in the
results, respectively. 
The uncertainty in the SPACAL
electromagnetic energy scale yields a 4\% systematic effect.
\item {\it Trigger Efficiency.} The trigger efficiency uncertainty results in a 7\% error.
\item {\it Radiative Corrections.} 
Radiative corrections have been 
estimated to result in a change in the cross-sections which is 
typically less than 5\%.
As no simulation of resolved photon processes which includes
radiative corrections is available, the estimate is based on direct events
only. The data are not corrected for this and it is not included in
the systematic error.
\item {\it Soft Underlying Event.} The unfolding procedure might also be influenced by 
the soft underlying event.
In Figure~\ref{fig:jetprof0}
we show the measured transverse energy flow about the two highest $E_t$ jets.
The flow is calculated from the energy clusters in a
strip of $\pm 1$ unit of $\eta$  with respect to the jet axes 
as a function of the difference between the $\phi$ of the clusters and the
$\phi$ of the jet axis.
Tracks are also included, modified as described in section 4.
The distribution is made for each of the two highest $E_t$ jets in the 
events and separated into different ranges of jet $E_t$ and $\eta$.
Because the
two jets are constrained to be close together in $\eta$, the energy flow
associated with the other approximately back-to-back jet is clearly visible. 
Note that the flow produced by this other jet is dependent on event topology 
as well as on the jet properties.
As the two jets are not precisely back-to-back, by orienting the flow in
$\phi$ such that the axis of the other 
jet is always to the left, we expose a wider, more clearly visible pedestal 
region to the right. 

The flow is compared with
the HERWIG(HO)/DG simulation with 10\% soft-underlying event and with predictions
from RAPGAP(HO)/SAS-2D which includes no model for the soft underlying event. Both
simulations give a good description of the energy flow in the core of the 
jets. 
Neither model is able to describe the energy flow in the pedestal region for all ranges of $E_t$ and
$\eta$. For $\eta < -2.0$, the pedestal is well-described by RAPGAP.
In the forward region, $\eta > -1.25$, the data lie between the RAPGAP and HERWIG
predictions. 
Although HERWIG overestimates
the data in this inclusive plot, we find that 10\% soft underlying event is needed to
account for the pedestal observed for $Q^2 \lsim 8\; {\rm GeV}^2$ (not shown separately)~\cite{marek}. 
The unfolding
was repeated using HERWIG(HO)/DG with 0\% and 15\% soft underlying event.
Differences in the resulting distributions were found to be 
$\sim 3\%$ and are included in the systematic errors. The jet pedestal has
more influence on the measurement of the effective parton density and will be
discussed again later in section 6.
\end{itemize}

\subsection{Discussion of the Results} 
The corrected triple-differential cross-section for $0.1 < y < 0.7$ and for jets
satisfying the criteria given in equations 10 -- 13 is given in Tables
~\ref{Tab2a} and~\ref{Tab2b}
and is shown in Figures \ref{fig:d3xg}, \ref{fig:d3et2} and \ref{fig:d3q2} 
in various projections. The three Figures show the cross-section
as functions of
$\xgjets$, $\etbarsq$ and  $Q^2$, respectively. In each case, the distributions
are shown for  
ranges of the other
two variables. The data are depicted as points with error
bars. The error bars show 
the statistical and systematic errors added in quadrature. Note that the systematic errors
are dominant everywhere. 
The absence of a data point indicates that no measurement was made in that
bin because of insufficient statistics for the unfolding.

In Figure~\ref{fig:d3xg}, the $\xgjets$ distributions can be seen to peak towards 
$\xgjets=1$, where the direct photon contribution is expected to be
concentrated. 
There is a strong decrease in the cross-section with increasing
$\etbarsq$. 
As $Q^2$ increases at fixed $\etbarsq$, the cross-section decreases 
and, for $\etbarsq < 150\;{\rm GeV}^2$, the relative contribution
from resolved photons, in the region $\xgjets \lsim
0.75$, can be seen to diminish. For  $\etbarsq > 150\;{\rm GeV}^2$, only 
the highest $\xgjets$ point has been measured.
Note that the reduction in the cross-section in the lowest $\xgjets$ bin
is a consequence of the cuts in $\overline{\eta}$ and $\etbarsq$. 

The data are compared with predictions of models with LO matrix elements and
parton densities. The 
HERWIG(LO)/DG simulation is shown
for two choices of the $Q^2$
suppression factor, $\omega$. The predictions for the direct only contribution
are also shown. With $\omega=0.1\;{\rm GeV}$, the HERWIG(LO)/DG model gives a
reasonable description
of the cross-section throughout the $Q^2$-$\etbarsq$ range. Increasing $\omega$
to 0.2 GeV leads to an overestimation in the low $Q^2$, low $\etbarsq$ regions.
The value of $\omega$ which best describes the data, however, depends on 
the frequency of soft underlying events. 
The 
RAPGAP(LO)/DG model prediction (not shown in this Figure), which contains 
no soft underlying event,  
requires $\omega$ $\sim 0.2\; {\rm GeV}$ in order to describe the data.  
As $Q^2$ increases, both HERWIG models tend to underestimate
the cross-section for intermediate $\xgjets$ values and overestimate it at high $\xgjets$
The direct only contribution is able to describe the data in the 
highest $\xgjets$ bin but 
underestimates the data for $\xgjets \lsim
0.75$ except possibly in the highest $Q^2$ range.

The $\etbarsq$ dependence of the cross-section is shown in
Figure~\ref{fig:d3et2}. 
It is compared
with predictions from RAPGAP(LO) using three different choices for the photon
parton densities. 
In the highest
$\xgjets$ range, where direct processes are expected to dominate, all the models
give a good description of the data. 
Elsewhere, the models provide a spread of predictions and no single
one is preferred.
The Drees-Godbole model tends to overestimate the cross-section 
for $1.6 < Q^2 < 3.5 \;{\rm GeV}^2$. 
For $Q^2 > 3.5 \; {\rm GeV^2}$ and 
$0.45 < x_\gamma < 0.75$, all three models tend to underestimate the data but are
still compatible within errors. 

The
$Q^2$ dependence of the cross-section is shown in Figure~\ref{fig:d3q2}. 
There is a steep decrease in the cross-section with $Q^2$. 
The expectation from the direct photon component only of the RAPGAP(LO)/DG model
shows a rate of suppression of the cross-section which 
is independent of  
$\xgjets$. However, the data 
show a rate of suppression which diminishes with increasing $\xgjets$.
This behaviour is governed by the additional $Q^2$ suppression of the photon
parton densities and 
the full RAPGAP(LO)/DG model including the resolved photon 
component gives a better description. 

From the above comparisons we conclude that the observed 
dependence of the dijet cross-section is
consistent with that predicted for a resolved virtual photon with parton
density functions evolving with $Q^2$ according to QCD motivated models. In
the next section we therefore proceed to extract an effective parton density
for virtual photons from the data.

\section{The effective parton density for virtual photons}
\subsection{Measurement of the Effective Parton Density}
In order to measure the parton densities of the 
virtual photons we adapt the Single Effective
Subprocess  
Approximation~\cite{SES}, originally developed for use in $p\overline{p}$
collisions and recently used to investigate 
real photon structure~\cite{Rick}.  In LO, the cross-section for the
production of dijets by resolved virtual photons can be
written as:
\begin{align}
& \frac{{\rm d}^5\sigma}{{\rm d}y \; {\rm d}x_{\gamma} \; 
                       {\rm d}x_{\rm p} \; {\rm d}\cos\theta^* \; {\rm
                       d}Q^2} = \qquad \qquad \qquad \notag  \\
& \qquad  \frac{1}{32 \pi s_{\rm ep}} 
\,\, \sum_{k=T,L}\frac{f_{\gamma/{\rm e}}^k(y,Q^2)}{y} 
\,\, \sum_{ij} 
\,\, \frac{f_{i/\gamma}^k(x_\gamma,P_t^2,Q^2)}{x_\gamma} 
\,\, \frac{f_{j/{\rm p}}(x_{\rm p},P_t^2)}{x_{\rm p}}  
\,\, |M_{ij}(\cos\theta^*)|^2  \; .
\label{eq:fullsig}
\end{align}
Here $f_{i/\gamma}^T$, $f_{i/\gamma}^L$ and $f_{i/{\rm p}}$ are the 
densities of parton species $i$ in transverse photons, longitudinal photons 
and the proton respectively. They are evaluated at the factorisation 
scale which we set equal to the renormalisation scale and choose to be
$P_t^2$. The $M_{ij}$ are matrix elements for 
$2\rightarrow 2$ parton-parton
hard scattering processes. The quantity $s_{{\rm ep}}$ is the square of the centre of mass
energy in the ep collision, $\theta^*$ is the polar angle of the outgoing
partons in the parton-parton centre of mass frame and $x_{\rm p}$ is the
momentum fraction of the parton out of the proton.
The fluxes of
transverse
and longitudinal photons are given by~\cite{Budnev}:
\begin{eqnarray}
 f^{T}_{\gamma/e}(y,Q^2) & = &
\frac{\alpha}{2\pi}\Bigl[\frac{1+(1-y)^2}{y}\frac{1}{Q^2}-\frac{2m_e^2y}{Q^4}\Bigr]
\label{eq:fT} \nonumber\\
f^{L}_{\gamma/e}(y,Q^2) & = &
\frac{\alpha}{2\pi}\frac{2(1-y)}{y}\frac{1}{Q^2}.
\label{eq:fL}
\end{eqnarray}
The measured dijet cross-sections do not allow us to separate 
the various parton sub-processes or photon 
polarisation states. To obtain a factorisable form, 
we first approximate equation~(\ref{eq:fullsig}) by:
\begin{align}
& \frac{{\rm d}^5\sigma}{{\rm d}y \; {\rm d}x_{\gamma} \; 
                       {\rm d}x_{\rm p} \; {\rm d}\cos\theta^* \; {\rm
                       d}Q^2} \approx \qquad \qquad \qquad \notag  \\
& \qquad  \frac{1}{32 \pi s_{\rm ep}} 
\,\, \frac{f_{\gamma/{\rm e}}^T(y,Q^2)}{y} 
\,\, \sum_{ij} 
\,\, \frac{f_{i/\gamma}(x_\gamma,P_t^2,Q^2)}{x_\gamma} 
\,\, \frac{f_{j/{\rm p}}(x_{\rm p},P_t^2)}{x_{\rm p}}  
\,\, |M_{ij}(\cos\theta^*)|^2  \; ,
\label{eq:redsig}
\end{align}
We have defined a set
of photon polarisation-averaged parton densities:
\begin{eqnarray}
f_{i/\gamma}(x_\gamma,P_t^2,Q^2) & \equiv &
f_{i/\gamma}^T(x_\gamma,P_t^2,Q^2)+\overline{\epsilon}f_{i/\gamma}^L(x_\gamma,P_t^2,Q^2),
\end{eqnarray}
where $\overline{\epsilon} \sim 1$ is the ratio of longitudinal to
transverse photon fluxes, averaged over the $y$-range of the data. 
$f_{i/\gamma}^L$ is expected to be small over most of the kinematic 
range considered here~\cite{vpth2,vpth6,frixetal}.

The Single Effective Subprocess (SES) approximation exploits the fact that
the dominant contributions to the cross-section, namely $qg \rightarrow qg$, 
$gg \rightarrow gg$ and $qq \rightarrow qq$ t-channel processes, 
come from parton-parton scattering matrix elements 
that have similar shapes and so differ mainly by
their associated colour factors. Thus the sum over processes can be
replaced by a single effective sub-process cross-section
and effective parton densities for the photon and proton:
\begin{align}
& \frac{{\rm d}^5\sigma}{{\rm d}y \; {\rm d}x_{\gamma} \; 
                       {\rm d}x_{\rm p} \; {\rm d}\cos\theta^* \; {\rm
                       d}Q^2} \approx \qquad \qquad \qquad \notag  \\
& \qquad  \frac{1}{32 \pi s_{\rm ep}} 
\,\, \frac{f_{\gamma/{\rm e}}^T(y,Q^2)}{y}  
\,\, \frac{\tilde{f}_{\gamma}(x_\gamma,P_t^2,Q^2)}{x_\gamma} 
\,\, \frac{\tilde{f}_{{\rm p}}(x_{\rm p},P_t^2)}{x_{\rm p}}  
\,\, |M_{\rm SES}(\cos\theta^*)|^2  \; .
\label{eq:SESsig}
\end{align}
where the effective parton densities are
\begin{eqnarray}
\tilde{f}_{\gamma}(x_\gamma, P_t^2,Q^2) & \equiv &
\sum_{\rm n_f} \left(f_{{\rm q}/\gamma}(x_\gamma, P_t^2,Q^2)
               +f_{\overline{\rm q}/\gamma}(x_\gamma, P_t^2,Q^2)\right) +
\frac{9}{4} \,  f_{{\rm g}/\gamma}(x_\gamma, P_t^2,Q^2)
\label{eq:efffgamma}
\;\;\;\;\\
{\rm and}\quad
\tilde{f}_{\rm p}(x_{\rm p}, P_t^2) & \equiv &
\sum_{\rm n_f} \left(f_{{\rm q}/{\rm p}}(x_{\rm p}, P_t^2)
               +f_{\overline{\rm q}/{\rm p}}(x_{\rm p}, P_t^2)\right) +
\frac{9}{4} \,  f_{{\rm g}/{\rm p}}(x_{\rm p}, P_t^2)
\; 
\end{eqnarray}
and the sums are over the quark flavours.

To extract the effective parton densities, 
the Bayesian unfolding method was applied again to 
correct the dijet cross-section to the diparton cross-section.
This second unfolding 
corrects for hadronisation effects, the influence of the soft underlying 
event, and initial and final state QCD radiation. It uses correlations between
the $\{x_{\gamma}^{jets},\overline{E}_T^2\}$ of the jets and $\{x_\gamma,
P_t^2\}$ of the parton-parton hard scattering, obtained using the
HERWIG(LO)-DG simulation. Here we use LO parton densities and the 1-loop
formula for $\alpha_S$ for a consistent leading-order treatment.
The resolution in $x_{\gamma}$ varies from $15\%$ at low $x_{\gamma}$ to
$10\%$ at high $x_{\gamma}$ in resolved events, and is $7\%$ in direct processes.
The $P_t^2$ resolution is $40\%$ at low $P_t$ improving to $24\%$ at high $P_t$.
The effective parton density was then determined by 
comparing the diparton cross-section measured in the data with that
predicted by the simulations with a known set of photon parton
densities:
\begin{equation}
\tilde{f}^{DATA}_\gamma = \tilde{f}^{MC}_\gamma \times \frac{(\tripart)^{DATA}}{(\tripart)^{MC}}
\end{equation} 
using equation~\ref{eq:efffgamma} to evaluate $\tilde{f}^{MC}_\gamma$. 

Now we discuss the systematic errors on the effective parton density.
Additional systematic errors associated with
the second unfolding were added in quadrature to those
associated with determination of the
triple-differential cross-section measurement described in the previous
section.  
The most significant new systematic effects 
arise from the model dependences.
These were estimated by repeating the second unfolding with RAPGAP and
HERWIG varying the input parton
densities, the amount of soft underlying event and the hadronisation models
as detailed below.
\begin{itemize}
\item {\it Model Dependence.} Additional model dependences arising from
the use of different parton showering and hadronisation
mechanisms and different input parton densities were 
estimated by unfolding the data using the RAPGAP(LO)/SAS-1D, 
RAPGAP(LO)/SAS-2D and RAPGAP(LO)/DG models
and comparing the results with those obtained after
unfolding with HERWIG(LO)/DG.
We assign a further 25\% systematic error on the 
basis of this test, of which 20\% arises from the hadronisation uncertainties. 
\item {\it Unfolding Instability.} 
Unfolding instabilities were estimated by varying the number of iterations
and lead to a 10\% uncertainty.
\item {\it Soft Underlying Event.}
The transformation from jet-based observables to parton variables is
more strongly influenced by the presence of the soft underlying event
than was the case for the transformation between true and
measured jets. As an independent measure of the amount of soft underlying event we
examine the transverse energy flow in the region outside the jets.
In Figure~\ref{fig:eout} we show the transverse energy
flow per unit area in the $(\eta,\phi)$ plane outside of circles of radius 1.3 about the two highest $E_t$
jets and in the range $-1.0 < \eta < 1.0$
as a function of $Q^2$. This central region of pseudo-rapidity is where
transverse energy flow from remnant-remnant interactions is expected to be largest.
The energy flow is corrected for detector effects. The inner error bars show
the statistical errors and the outer error bars 
the quadratic sum of statistical and systematic errors. The
latter include all the sources considered for the jet cross-section measurement.
The observed decrease with increasing $Q^2$ is compared with predictions from
HERWIG(LO)/DG with 0\%, 5\% and 10\% soft underlying event and with RAPGAP(LO)/SAS-2D.
Note that with the
different choice of minimum $P_t$ and space-like shower parameter QSPAC (see
section 3), which is used in HERWIG(LO)-DG
for this step in the analysis, a lower soft underlying event frequency is
required to describe the data.

We varied the soft-underlying event probability in HERWIG between 0\%
and 10\% in order to estimate the systematic uncertainty on the effective 
parton density. The resulting systematic error 
has a mean value of
20\% and the error is largest at low $x_{\gamma}$ and low
$Q^2$.
\end{itemize}
\subsection{Discussion of the Extracted Effective Parton Density}
The extracted effective parton density is given in Table~\ref{Tab3} and shown in Figures~\ref{fig:epdfxg}, \ref{fig:epdfpt2}, and
\ref{fig:epdfq2}. The measurement uncertainties are everywhere domininated by
the systematic errors. Only points where $\langle P_t^2\rangle > \langle Q^2\rangle$ are shown. It should be 
noted that in the highest $Q^2$ bins, $Q^2$ approaches $P_t^2$ 
and the assumptions of factorisation involved in the definition of
a universal effective parton density begin to break down~\cite{vpth2, SAS}. Furthermore, the effects arising from 
higher twist contributions and longitudinally polarised photons are largest
here. 
Nevertheless, we extract the effective parton distributions in the 
full region of the available data. The universality of the PDF's 
extracted in the region where $Q^2$ is of the same order as $P_t^2$
will need to be demonstrated in other virtual photon induced reactions.
The Figures show the evolution of the effective parton density in the photon both with
the scale at which it is probed and with its virtuality. 
The data are compared with three sets of parton
densities, SAS-1D, SAS-2D and the effective parton density calculated from
the Drees-Godbole model using GRV-LO densities for the real photon and
setting the free parameter, $\omega$, in the suppression factor to 0.1 GeV. 
The predictions were evaluated at the mean $x_\gamma$ and logarithmic mean
$Q^2$ and $P_t^2$ of the ranges.

Figure~\ref{fig:epdfxg} shows that the effective parton density tends to rise 
with $x_\gamma$ in the region studied. This shape is characteristic of photon
structure.
The data are described by all three models within
errors except possibly in the two higher $Q^2$ ranges where the models tend to
underestimate the data in the intermediate and high $x_{\gamma}$ bins.

In Figure~\ref{fig:epdfpt2}, the parton density is shown
as a function of the square of the probing scale $P_t^2$ 
in ranges of $x_\gamma$ and $Q^2$. 
The effective parton density is roughly independent of $P_t^2$ in each of the
ranges. This scale dependence contrasts with the falling behaviour expected
for hadronic parton densities and, within the rather large systematic errors, is
consistent with the normalisation and logarithmic scale dependence predicted 
by perturbative QCD for the anomalous component of the photon. 

The decrease of the parton density with virtuality is
most clearly seen in Fig~\ref{fig:epdfq2}, where the $Q^2$ dependence is shown in ranges of
$P_t^2$ and $x_\gamma$. 
The three parameterisations for the parton density all give a good description of
the data both in the lowest $x_{\gamma}$ range and in the lowest two $Q^2$ bins. 
They all predict a
more rapid suppression as $Q^2$ increases 
than is seen in the
data. It is in this region where $Q^2 \rightarrow P_t^2$ that 
non-leading terms, not accounted for in the
models, are expected to become important and may affect the 
extraction of the effective parton distribution from these data.

In Figure~\ref{fig:epdfq22}, the $Q^2$ evolution of the effective parton density is shown for
$P_t^2=85\; {\rm GeV}^2$, in the upper Figure
for
$0.35 < x_\gamma < 0.5$ and in the lower for $0.5 < x_\gamma < 0.7$.
Superimposed on the same plots are photoproduction data at $P_t^2 =112\; {\rm GeV}^2$ 
and $x_\gamma$ values of 0.3 and 0.55, taken from reference~\cite{Rick}, 
which we have extrapolated to the $P_t^2$ and
$x_\gamma$ values in the Figures. The extrapolation was based on
the GRV-LO parton densities which give a good description of the photoproduction data. The evolution is
compared with that predicted by the three models described above and also with a
simple $\rho$-pole suppression factor characteristic of a pure VMD model:
\begin{equation}
f_{i|\gamma}(x_\gamma, P_t^2, Q^2) \sim f_{i|\gamma}(x_\gamma, P_t^2, 0)\Bigl(\frac{m_{\rho}^2}{m_{\rho}^2+Q^2}\Bigr)^2.
\label{eq:rhopole}
\end{equation}
The latter 
clearly underestimates the data. The logarithmic suppression predicted by the
virtual photon models on the other hand gives a good description below $Q^2
\sim 20\; {\rm GeV}^2$.
At higher $Q^2$ they predict a more rapid decrease than is seen in the data.

\section{Conclusions}
We have measured the triple-differential cross-section, $\triple$, in dijet
events for $1.6 < Q^2 < 80\; \rm{GeV}^2$ and $0.1 < y < 0.7$. 
The measured
cross-sections show the $\xgjets$, $\etbarsq$ and $Q^2$ behaviour expected for
processes in which a virtual photon, carrying a partonic structure evolving
according to perturbative QCD, interacts with the proton via hard
parton-parton scattering.
The measurements are consistent with the perturbative QCD prediction 
that, as $Q^2 \rightarrow \etbarsq$, the photon structure
reduces to a simple direct coupling to $q\overline{q}$ pairs and the dijet
cross-section is well-described without invoking photon structure. LO Monte
Carlo models based on these QCD predictions give a good description of the data.

An effective, leading order, parton density of the virtual photon has been
extracted in the Single Effective Subprocess approximation and its dependences
on $x_\gamma$, probing scale $P_t^2$ and target virtuality $Q^2$ have been
measured. The photon parton density is approximately independent of $p_T^2$
and, within errors, it is consistent with the normalisation and 
logarithmic scaling violations 
characteristic of 
photon structure. It is seen to be suppressed with
increasing $Q^2$ as predicted by perturbative QCD. 

\section*{Acknowledgements}
We are grateful to the HERA machine group whose outstanding
efforts have made and continue to make this experiment possible. 
We thank
the engineers and technicians for their work in constructing and now
maintaining the H1 detector, our funding agencies for 
financial support, the
DESY technical staff for continual assistance, 
and the DESY directorate for the
hospitality which they extend to the non-DESY 
members of the collaboration. 
We wish to thank B.\ P\"{o}tter for many helpful
discussions.


\newpage
%
%
%
%
%
%
%
%
%
\begin{table}
\begin{center}
\begin{tabular}{|c|c|c|c|c|c|c|}
  \hline 
  $Q^2$ (${\rm GeV}^2)$ & $\etbarsq$ (${\rm GeV}^2$) & $x_\gamma$ & $\sigma$
  (pb) & Stat($\pm$) & Sys(+) & Sys(-)\ \\
  \hline\hline
1.6$<Q^2<$3.5 & 30$<\etbarsq<$45 &0.15$<x_\gamma<$0.3 &     7.50  &     0.18  &
   2.84    &    1.96  \ \\
 & &0.3$<x_\gamma<$0.45 &     5.98  &     0.15  &      1.64  &       1.37  \ \\
 & &0.45$<x_\gamma<$0.6 &     8.34  &     0.21  &      4.20  &      4.15  \ \\
 & &0.6$<x_\gamma<$0.75 &    12.00  &     0.25  &      2.59  &      2.58  \ \\
 &  &0.75$<x_\gamma<$1.0 &    18.76  &     0.30  &      1.70 &      1.71  \
  \\
\hline
1.6$<Q^2<$3.5 & 45$<\etbarsq<$65 &0.15$<x_\gamma<$0.3 &     2.81  &     0.11
  &      1.28    &    0.78  \ \\
 & &0.3$<x_\gamma<$0.45 &     2.20  &     0.09  &      0.85   &     0.70  \ \\
 & &0.45$<x_\gamma<$0.6 &     2.62  &     0.08  &      0.78   &     0.75  \ \\
 & &0.6$<x_\gamma<$0.75 &     4.58  &     0.11  &      1.21   &     0.97  \ \\
 &  &0.75$<x_\gamma<$1.0 &     9.22  &     0.19  &      1.38  &      1.51  \
  \\
\hline
1.6$<Q^2<$3.5 & 65$<\etbarsq<$150 &0.3$<x_\gamma<$0.45 &     0.33  &     0.02
  &      0.22   &     0.19  \ \\
 & &0.45$<x_\gamma<$0.6 &     0.58  &     0.03  &      0.14   &     0.15  \ \\
 & &0.6$<x_\gamma<$0.75 &     0.81  &     0.03  &      0.18   &     0.16  \ \\
 &  &0.75$<x_\gamma<$1.0 &     2.40  &     0.05  &      0.46  &      0.41  \
  \\
\hline
 1.6$<Q^2<$3.5 & 150$<\etbarsq<$300 &0.75$<x_\gamma<$1.0 &     0.27  &
  0.01  &      0.05   &     0.04  \
  \\
\hline\hline
3.5$<Q^2<$8.0 & 30$<\etbarsq<$45 &0.15$<x_\gamma<$0.3 &     1.88  &     0.05
  &      1.33    &    1.33  \ \\
 & &0.3$<x_\gamma<$0.45 &     1.95  &     0.05  &      0.47   &     0.30  \ \\
 & &0.45$<x_\gamma<$0.6 &     3.46  &     0.07  &      1.42   &     1.45  \ \\
 & &0.6$<x_\gamma<$0.75 &     5.18  &     0.09  &      1.06   &     1.07  \ \\
 &  &0.75$<x_\gamma<$1.0 &     7.38  &     0.12  &      0.68  &      0.64  \
  \\
\hline
3.5$<Q^2<$8.0 & 45$<\etbarsq<$65 &0.15$<x_\gamma<$0.3 &     0.75  &     0.03
  &      0.36   &     0.29  \ \\
 & &0.3$<x_\gamma<$0.45 &     0.81  &     0.02  &      0.22    &    0.18  \ \\
 & &0.45$<x_\gamma<$0.6 &     1.27  &     0.04  &      0.31    &    0.29  \ \\
 & &0.6$<x_\gamma<$0.75 &     1.72  &     0.05  &      0.43    &    0.46  \ \\
 &  &0.75$<x_\gamma<$1.0 &     3.80  &     0.07  &      0.46    &    0.43  \
  \\
\hline
3.5$<Q^2<$8.0 & 65$<\etbarsq<$150 &0.3$<x_\gamma<$0.45 &     0.21  &     0.01
  &      0.05    &    0.04  \ \\
 & &0.45$<x_\gamma<$0.6 &     0.21  &     0.01  &      0.06    &     0.06  \ \\
 & &0.6$<x_\gamma<$0.75 &     0.34  &     0.01  &      0.05    &    0.06  \ \\
 &  &0.75$<x_\gamma<$1.0 &     0.98  &     0.02  &      0.14    &    0.10  \
  \\
\hline
3.5$<Q^2<$8.0 & 150$<\etbarsq<$300  &0.75$<x_\gamma<$1.0 &     0.11  &
  0.01  &      0.02    &    0.02  \
  \\
\hline\hline
\end{tabular}


\end{center}
\caption{The triple-differential dijet cross-section 
\triple for $0.1 < y < 0.7$ in ranges of $Q^2$, $\etbarsq$ and $\xgjets$. The cross-section in pb
is given together with the statistical, positive systematic and negative
systematic errors.
The table shows measurements for 
$Q^2\leq 8.0~{\rm GeV}^2$. The higher $Q^2$ region is given in
table~\ref{Tab2b}.}
\label{Tab2a}
\end{table}
\begin{table}
\begin{center}
\begin{tabular}{|c|c|c|c|c|c|c|}
  \hline 
  $Q^2$ (${\rm GeV}^2$) & $\etbarsq$ (${\rm GeV}^2$) & $x_\gamma$ & $\sigma$
  (pb) & Stat($\pm$) & Sys(+) & Sys(-)\ \\
  \hline\hline
8.0$<Q^2<$25 & 30$<\etbarsq<$45 &0.15$<x_\gamma<$0.3 &     0.27  &     0.01
  &      0.21    &    0.20  \ \\
 & &0.3$<x_\gamma<$0.45 &     0.670 &     0.02  &      0.17    &    0.18  \ \\
 & &0.45$<x_\gamma<$0.6 &     0.670 &     0.01  &      0.16    &     0.14   \ \\
 & &0.6$<x_\gamma<$0.75 &     1.58  &     0.03  &      0.35    &    0.34  \ \\
 &  &0.75$<x_\gamma<$1.0 &     2.13  &     0.03  &      0.41   &     0.38  \
  \\
\hline
8.0$<Q^2<$25 & 45$<\etbarsq<$65 &0.15$<x_\gamma<$0.3 &     0.11  &     0.01
  &      0.05    &    0.04  \ \\
 & &0.3$<x_\gamma<$0.45 &     0.23  &     0.01  &      0.04   &     0.04  \ \\
 & &0.45$<x_\gamma<$0.6 &     0.31  &     0.01  &      0.08   &     0.08  \ \\
 & &0.6$<x_\gamma<$0.75 &     0.54  &     0.01  &      0.12   &     0.13  \ \\
 &  &0.75$<x_\gamma<$1.0 &     1.09  &     0.02  &      0.15   &     0.15  \
  \\
\hline
 8.0$<Q^2<$25 & 65$<\etbarsq<$150 &0.3$<x_\gamma<$0.45 &     0.047 &
  0.002 &      0.01   &     0.011 \ \\
 & &0.45$<x_\gamma<$0.6 &     0.074 &     0.003 &      0.022   &    0.020 \ \\
 & &0.6$<x_\gamma<$0.75 &     0.091 &     0.003 &      0.014   &    0.012 \ \\
 &  &0.75$<x_\gamma<$1.0 &     0.298 &     0.005 &      0.031   &    0.038 \
  \\
\hline
 8.0$<Q^2<$25 & 150$<\etbarsq<$300  &0.75$<x_\gamma<$1.0 &     0.039 &
  0.001 &      0.008   &    0.006 \
  \\
\hline\hline
25$<Q^2<$80 & 30$<\etbarsq<$45 &0.15$<x_\gamma<$0.3 &     0.021 &     0.001 &
  0.033   &    0.033 \ \\
 & &0.3$<x_\gamma<$0.45 &     0.065 &     0.005 &      0.019   &    0.017 \ \\
 & &0.45$<x_\gamma<$0.6 &     0.114 &     0.006 &      0.025   &    0.026 \ \\
 & &0.6$<x_\gamma<$0.75 &     0.236 &     0.007 &      0.081   &    0.078 \ \\
 &  &0.75$<x_\gamma<$1.0 &     0.433 &     0.007 &      0.087   &    0.086 \
  \\
\hline
25$<Q^2<$80 & 45$<\etbarsq<$65 &0.15$<x_\gamma<$0.3 &     0.018 &     0.001 &
  0.018   &    0.018 \ \\
 & &0.3$<x_\gamma<$0.45 &     0.055 &     0.005 &      0.026   &    0.036 \ \\
 & &0.45$<x_\gamma<$0.6 &     0.051 &     0.003 &      0.015   &    0.017 \ \\
 & &0.6$<x_\gamma<$0.75 &     0.120 &     0.005 &      0.040   &    0.040 \ \\
 &  &0.75$<x_\gamma<$1.0 &     0.220 &     0.004 &      0.031   &    0.020 \
  \\
\hline
 25$<Q^2<$80 & 65$<\etbarsq<$150 & 0.3$<x_\gamma<$0.45 &     0.010 &     0.001
  &      0.005   &    0.005 \ \\
 & &0.45$<x_\gamma<$0.6 &     0.024 &     0.001 &      0.006   &    0.009 \ \\
 & &0.6$<x_\gamma<$0.75 &     0.020 &     0.001 &      0.005   &    0.005 \ \\
 &  &0.75$<x_\gamma<$1.0 &     0.060 &     0.001 &      0.008   &    0.004 \
  \\
\hline
 25$<Q^2<$80 & 150$<\etbarsq<$300 &0.75$<x_\gamma<$1.0 &     0.012 &     0.001
  &      0.001   &    0.002 \
  \\
\hline\hline
\end{tabular}


\end{center}
\caption{The triple-differential dijet cross-section 
\triple in ranges of $Q^2$, $\etbarsq$ and $\xgjets$. The cross-section in pb
is given together with the statistical, positive systematic and negative
systematic errors.
The table shows measurements for 
$Q^2> 8.0~{\rm GeV}^2$. The lower $Q^2$ region is given in
table~\ref{Tab2a}.}
\label{Tab2b}
\end{table}
\begin{table}
\begin{center}
\begin{tabular}{|c|c|c|c|c|c|c|}
  \hline 
  $Q^2$ (${\rm GeV}^2$) & $P_t^2$ (${\rm GeV}^2$) & $x_\gamma$ & $\alpha^{-1}
  \tilde{f}_{\gamma}$ & Stat($\pm$) & Sys(+) & Sys(-) \ \\
  \hline\hline
2.4 & 40.0 & 0.275 &     0.55   &    0.02   &     0.23   &    0.19 \ \\
 & & 0.425 &     0.60   &    0.02   &     0.15   &    0.12 \ \\
 & & 0.6 &     0.95   &    0.03   &     0.17   &    0.29 \ \\
\hline
 & 52.0 & 0.275 &     0.59   &    0.02   &     0.32  &     0.19 \ \\
 & & 0.425 &     0.57   &    0.02   &     0.20    &   0.15 \ \\
 & & 0.6 &     0.93   &    0.03   &     0.19    &   0.18 \ \\
\hline
 & 85.0 & 0.425 &     0.53   &    0.02   &     0.29  &     0.18 \ \\
 & & 0.6 &     0.98   &    0.03   &     0.21   &    0.26 \ \\
\hline\hline
5.3 & 40.0 & 0.275 &     0.33   &    0.01   &     0.15   &    0.15 \ \\
 & & 0.425 &     0.49   &    0.02   &     0.14   &    0.13 \ \\
 & & 0.6 &     0.82   &    0.03   &     0.13   &    0.30 \ \\
\hline
 & 52.0 & 0.275 &     0.36   &    0.01   &     0.20   &    0.16 \ \\
 & & 0.425 &     0.50   &    0.02   &     0.14   &    0.15 \ \\
 & & 0.6 &     0.85   &    0.03   &     0.14  &     0.25 \ \\
\hline
 & 85.0 & 0.425 &     0.64   &    0.02   &     0.19   &    0.21 \ \\
 & & 0.6 &     0.87   &    0.03   &     0.15   &    0.23 \ \\
\hline\hline
12.7 & 40.0 & 0.275 &     0.16   &    0.01   &     0.07   &    0.08 \ \\
 & & 0.425 &     0.42   &    0.02   &     0.09   &    0.19 \ \\
 & & 0.6 &     0.54   &    0.02   &     0.10   &    0.25 \ \\
\hline
 & 52.0 & 0.275 &     0.18   &    0.01   &     0.09   &    0.09 \ \\
 & & 0.425 &     0.45   &    0.02   &     0.09   &    0.20 \ \\
 & & 0.6 &     0.62   &    0.03   &     0.10   &    0.23 \ \\
\hline
 & 85.0 & 0.425 &     0.46   &    0.02   &     0.12   &    0.21 \ \\
 & & 0.6 &     0.74   &    0.03   &     0.14   &    0.27 \ \\
\hline\hline
 40.0 & 85.0 & 0.425 &     0.43   &    0.04   &     0.09    &   0.27 \ \\
 & & 0.6 &     0.65   &    0.04   &     0.19    &   0.41 \ \\
\hline\hline
\end{tabular}

\end{center}
\caption{The leading order effective parton density of the photon 
$\xgamma \tilde{f}_{\gamma} = 
\sum_{\rm n_f} \left(f_{{\rm q}/\gamma}
               +f_{\overline{\rm q}/\gamma}\right) +
\frac{9}{4} \,  f_{{\rm g}/\gamma}
$
, divided by the fine structure constant $\alpha$,
for different values of $Q^2$, $P_t^2$ and $x_\gamma$. 
Statistical, positive systematic and negative systematic errors are given.}
\label{Tab3}
\end{table}
\begin{figure}[ht]
\begin{center}
\epsfig{file=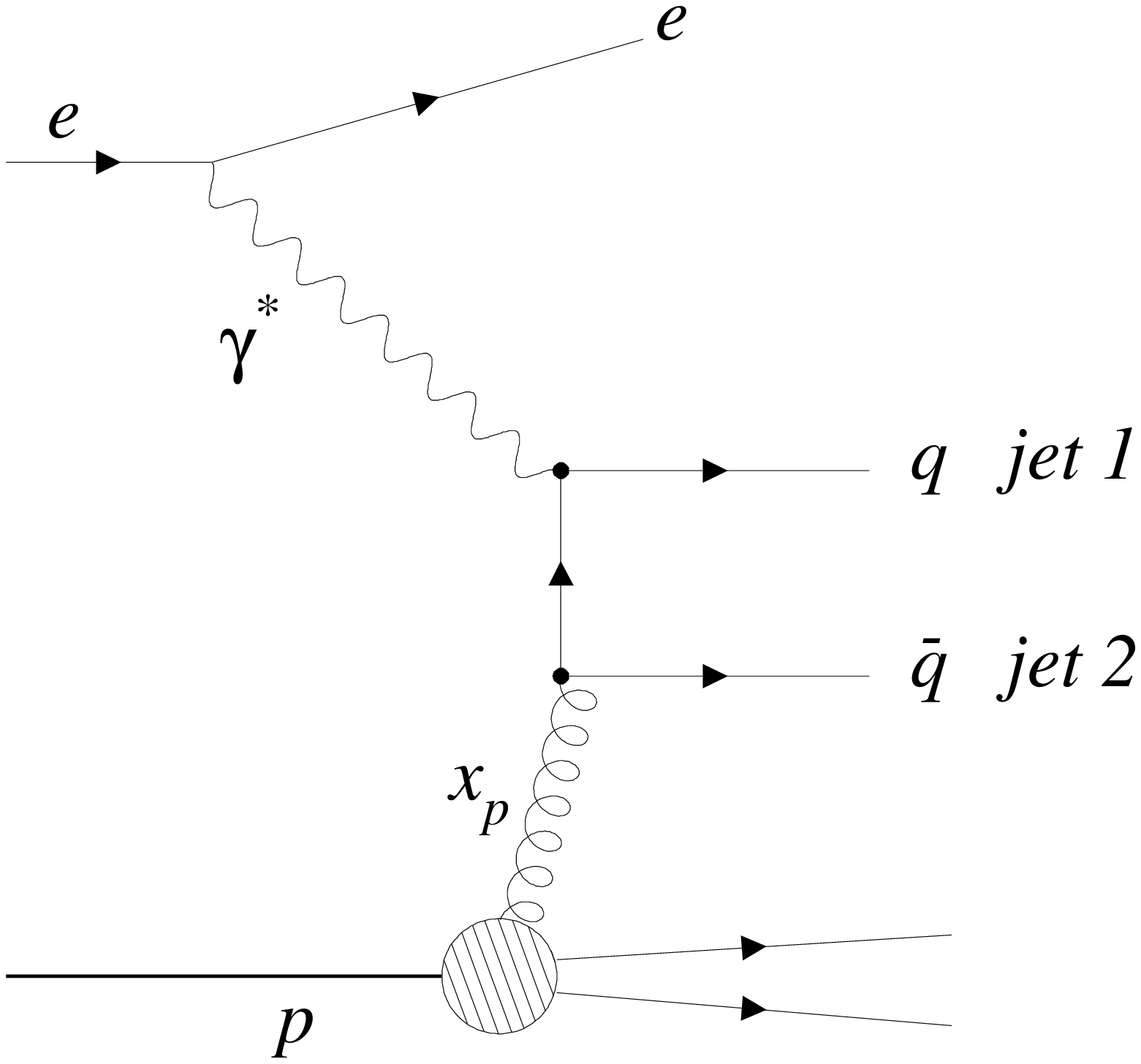,width=0.45\textwidth}
\epsfig{file=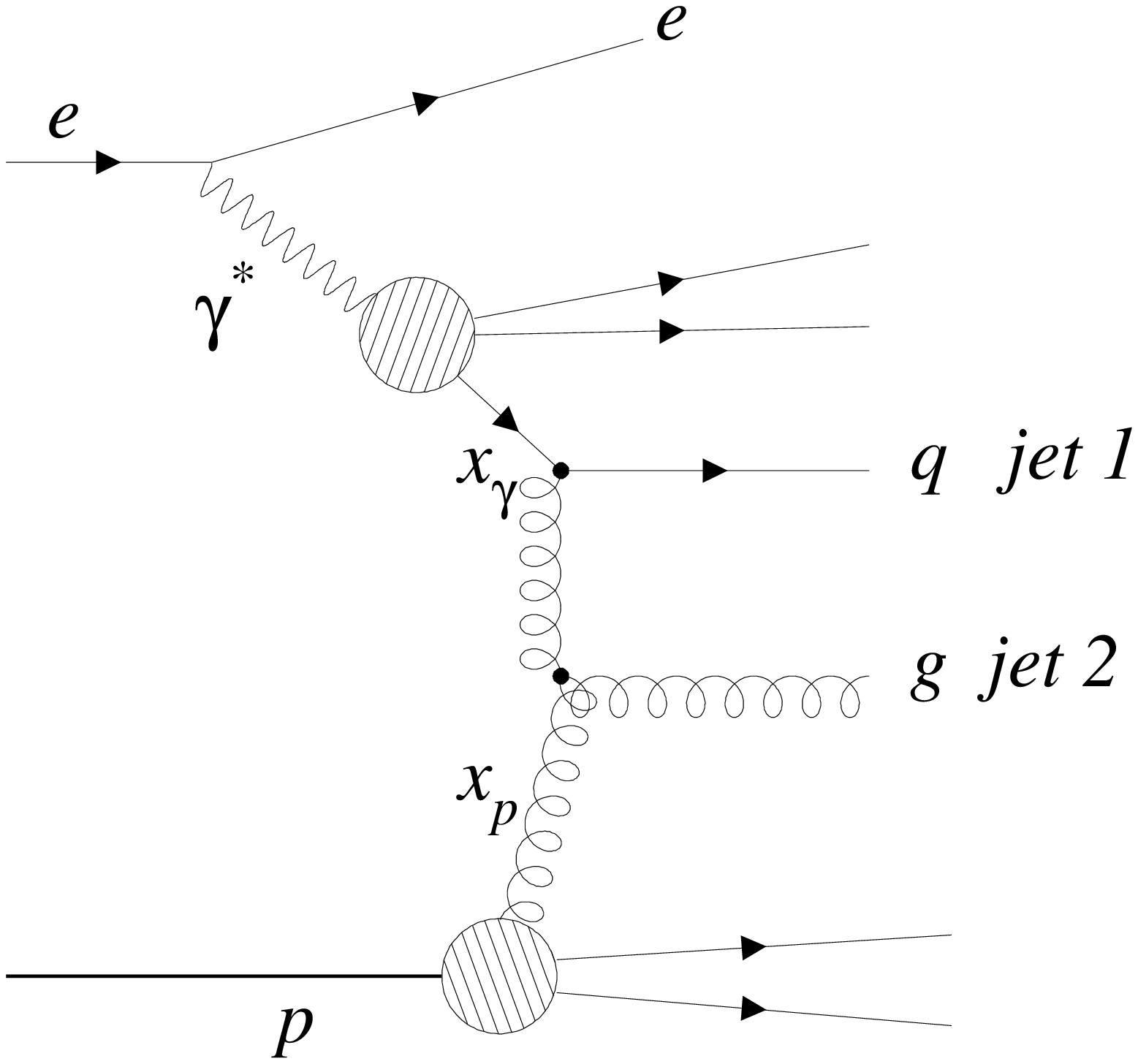,width=0.45\textwidth}
\caption{
Examples of leading order Feynman diagrams of dijet production in $\gamma^* p$ collisions. 
In direct processes
(left) all of the photon's 4-momentum enters the hard collision. If
the transverse momentum of the outgoing dijets (labelled jet 1 and jet 2 in the
figures) is large compared with $Q^2$,
the virtual photon may be resolved (right) with only a fraction $x_{\gamma} <
1$ of the photon's 4-momentum involved. The diagrams are applicable to
dijet production with either real or virtual photons.}
\label{fig:dirres}
\end{center} 
\end{figure}
\begin{figure}[ht]
\begin{center}
\epsfig{file=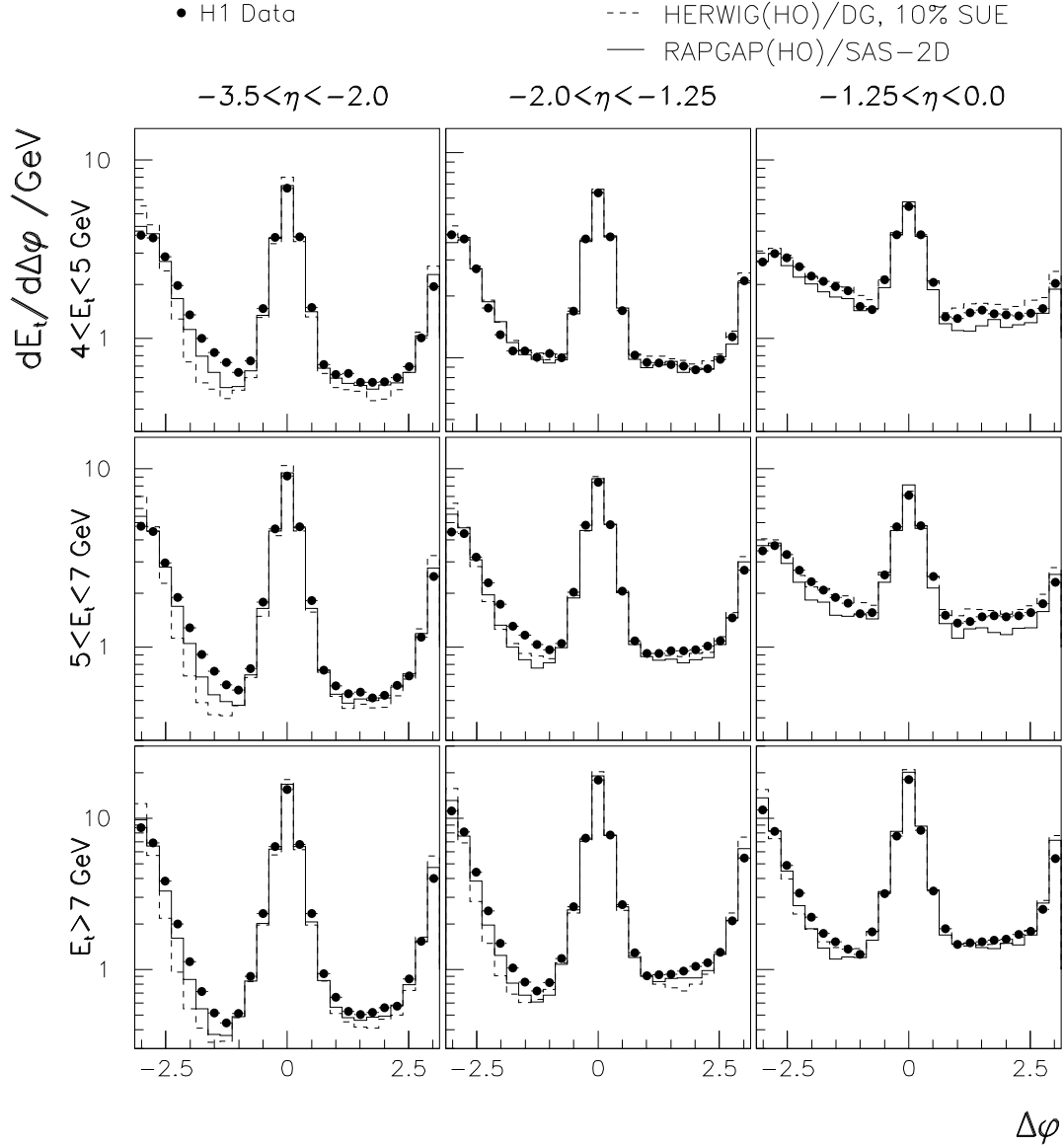,width=\textwidth}
\caption{
The observed transverse energy flow in $\phi$ 
with respect to the axes (located at $\phi = 0$) of each
of the two highest $E_t$ jets in the selected event sample for a 
slice $\mid\eta-\eta^{jet}\mid < 1$. The result is shown for various
ranges of $\eta$ and $E_t$ of the jets. The second jet is always
chosen to have $\phi<0$, leaving the pedestal level clearly
visible to the right of the jet core, in the region $\phi>0$. 
The data are compared with    
predictions from the HERWIG(HO)/DG simulation
with 10\% soft 
underlying event (dashed histogram) and with the RAPGAP(HO)/SAS-2D
model.}
\label{fig:jetprof0}
\end{center} 
\end{figure}
\newpage
\begin{figure}[ht]
\begin{center}
\epsfig{file=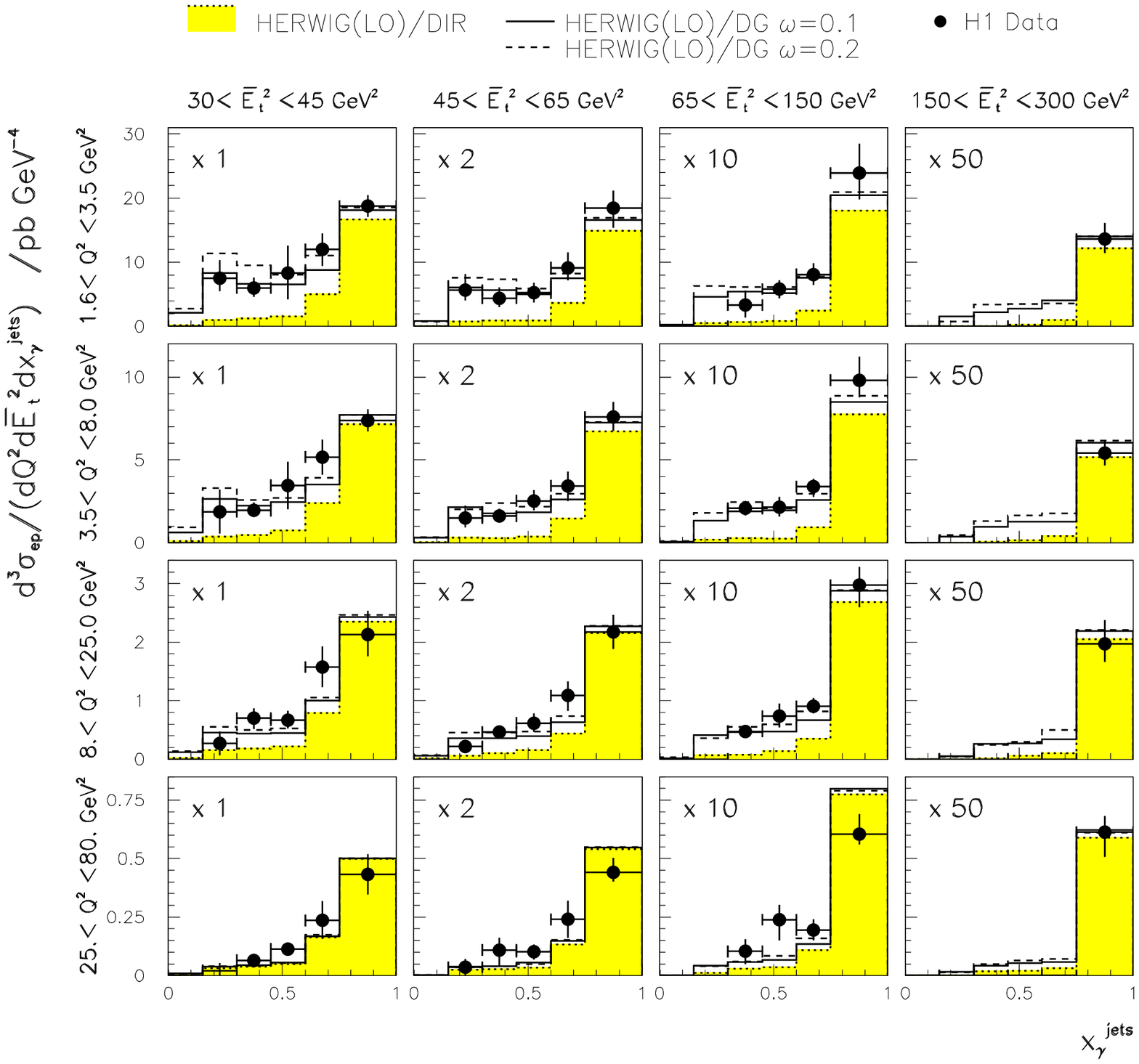,width=\textwidth}
\caption{The differential dijet cross-section 
\triple shown as a function of $\xgjets$ for different regions of $\etbarsq$ and
$Q^2$. The scale factors applied to the cross-sections are
indicated.
The  error
bar shows the quadratic sum of systematic and statistical errors.
The absence of a data point indicates that no measurement was made because
of insufficient statistics for the two dimensional unfolding.
Also shown is the HERWIG(LO)/DG model
with 10\% 
soft-underlying event and two choices of the $Q^2$ suppression
factor $\omega$. The full histogram is for $\omega=0.1~{\rm GeV}$ 
and the dashed for $\omega=0.2~{\rm GeV}$. The direct component 
of this model is shown as the shaded histogram.}
\label{fig:d3xg}
\end{center} 
\end{figure}
\newpage
\begin{figure}[ht]
\begin{center}
\epsfig{file=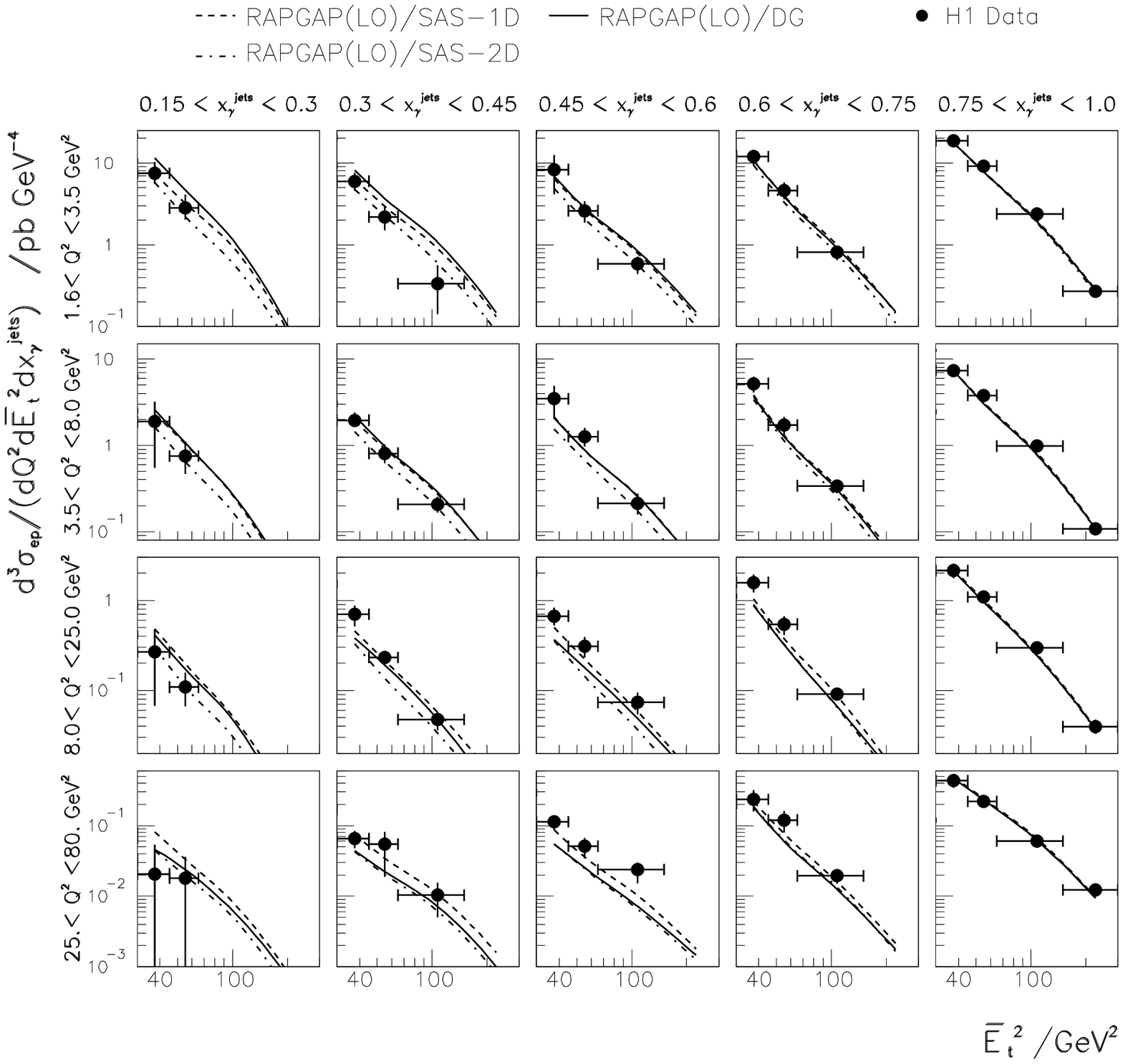,width=\textwidth}
\caption{The differential dijet cross-section 
\triple shown as a function of $\etbarsq$ for different regions of $\xgjets$ and
$Q^2$. 
The error
bars show the quadratic sum of systematic and statistical errors.
Also shown is the
RAPGAP(LO) model with three
choices of photon parton density. 
The DG model with GRV-LO real photon densities
and $\omega=0.2~{\rm GeV}$ is shown as the full curve. 
The predictions with SAS-1D and SAS-2D
are shown as the dashed and dot-dashed curves respectively.}
\label{fig:d3et2}
\end{center} 
\end{figure}
\newpage
\begin{figure}[ht]
\begin{center}
\epsfig{file=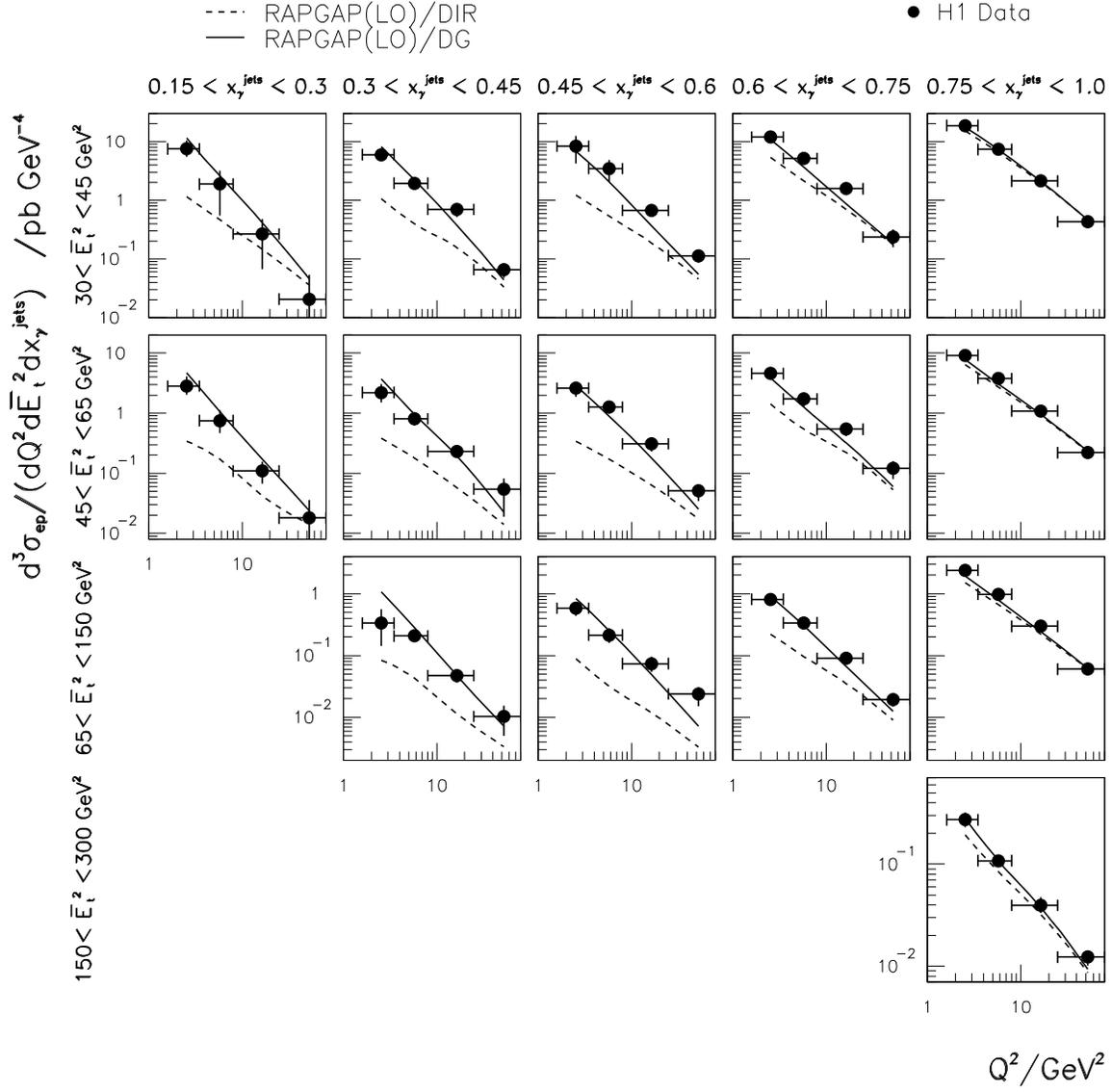,width=\textwidth}
\caption{The differential dijet cross-section 
\triple shown as a function of $Q^2$
for different regions of $\xgjets$ and $\etbarsq$.
The error
bars show the quadratic sum of systematic and statistical errors.
The prediction from the 
RAPGAP(LO)/DG model is shown as the full curve.
The dashed curve shows the 
prediction for the direct photon processes in this model.}
\label{fig:d3q2}
\end{center} 
\end{figure}
\newpage
\begin{figure}[ht]
\begin{center}
\epsfig{file=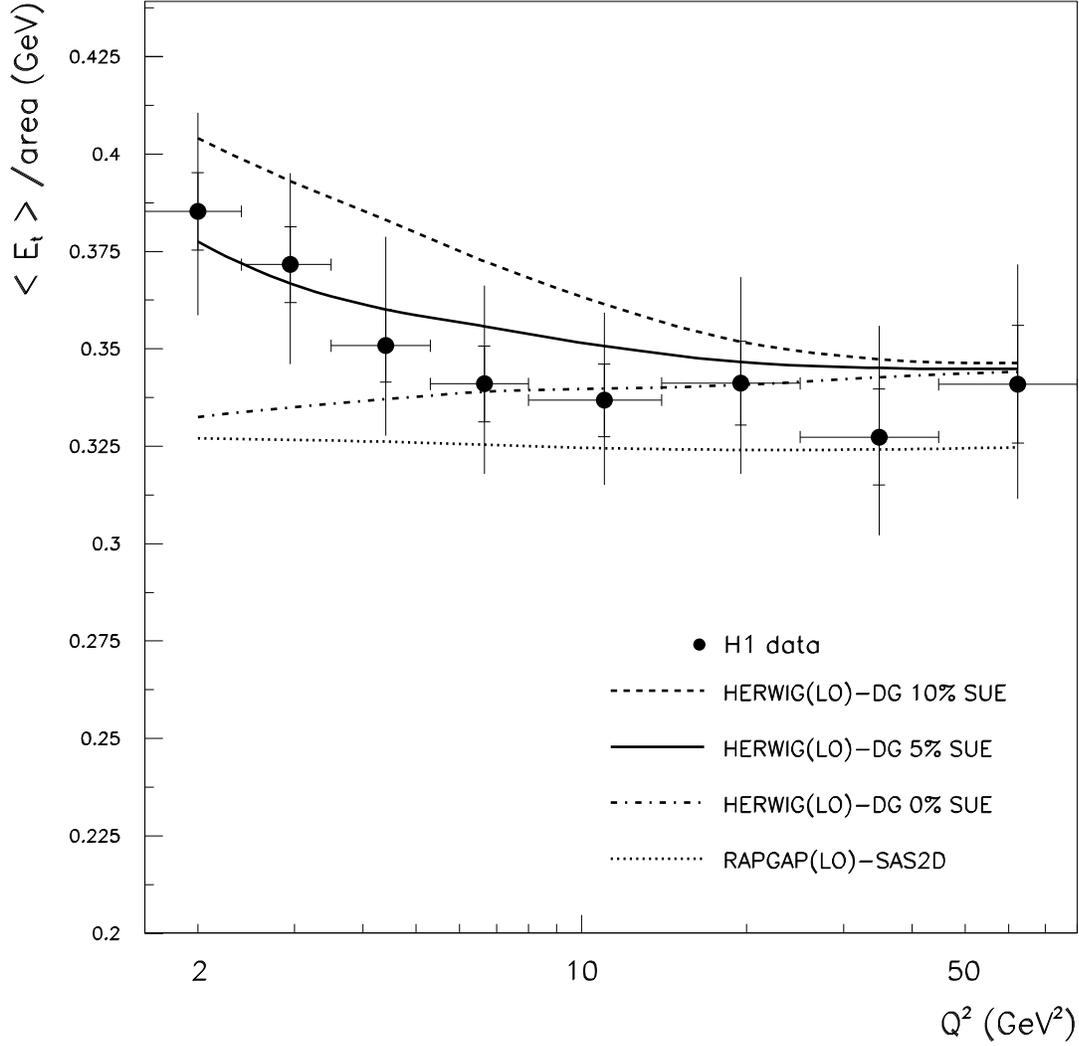,width=\textwidth}
\caption{The average transverse energy per 
unit area in $\eta - \phi$ space 
in the range $-1.0< \eta < 1.0$ and $0 < \phi < 2\pi$ in the $\gamma^{*}p$
cms as a function of $Q^2$ and outside the two highest $E_t$ jets. 
The data are corrected for detector effects. The inner error bars show the 
statistical errors and the outer error bars are the quadratic sum of the
statistical and systematic errors. Also shown is 
the prediction from the HERWIG(LO)/DG model
with three choices for
the percentage of resolved events with a soft underlying event:
0\% (dash-dotted), 
5\% (full) and 10\% (dashed). 
The prediction from the RAPGAP(LO)/SAS-2D model
is shown as the dotted curve.}
\label{fig:eout}
\end{center} 
\end{figure}
\newpage
\begin{figure}[ht]
\begin{center}
\epsfig{file=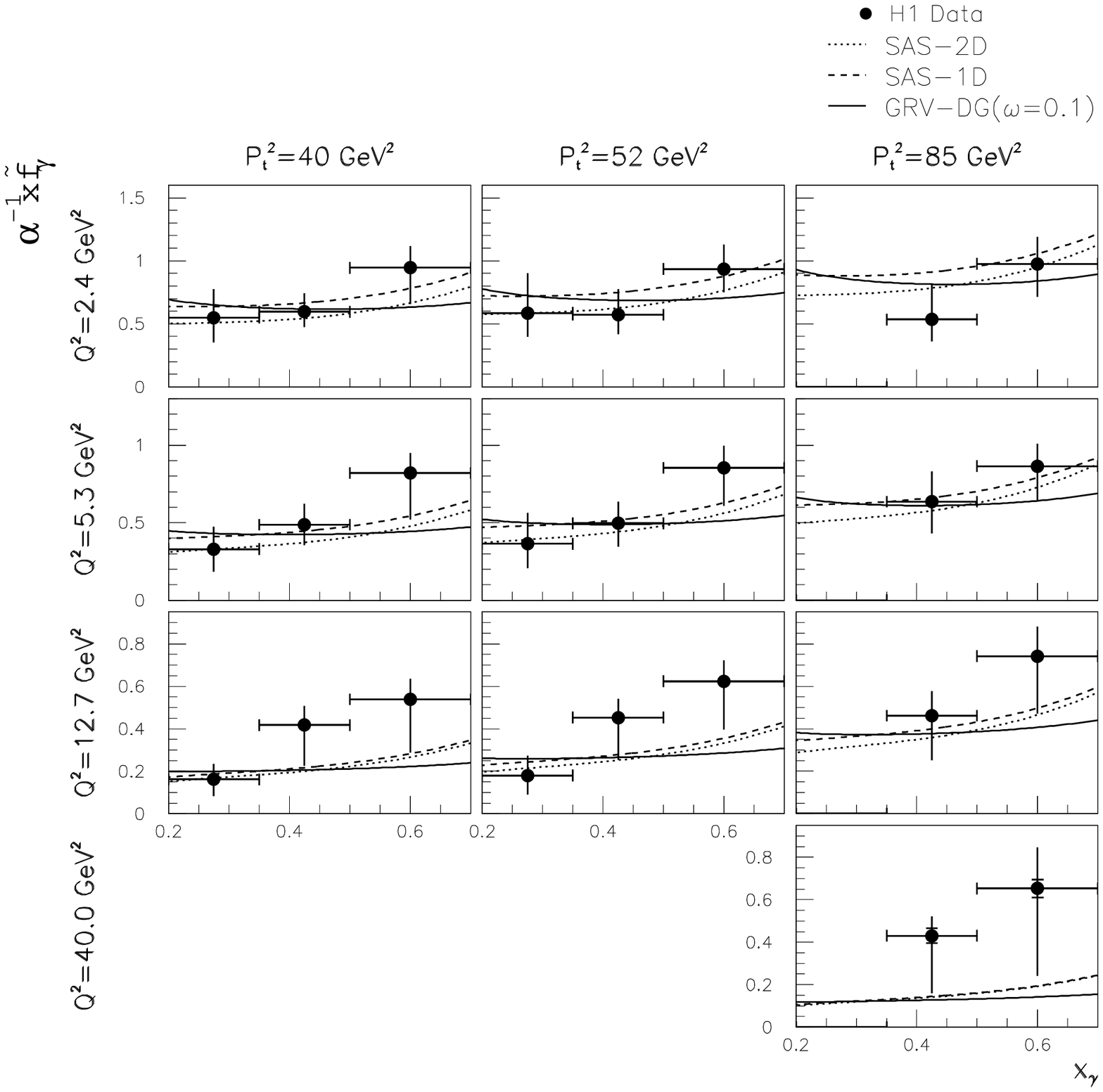,width=\textwidth}
\caption{The leading order effective parton density of the photon
$\xgamma \tilde{f}_{\gamma} = 
\sum_{\rm n_f} \left(f_{{\rm q}/\gamma}
               +f_{\overline{\rm q}/\gamma}\right) +
\frac{9}{4} \,  f_{{\rm g}/\gamma}
$
, divided by the fine structure constant $\alpha$,
as a function of $x_\gamma$ 
for different values of $Q^2$ and $P_t^2$. 
The data are displayed as points, with the inner error bar depicting
the statistical error, and the total error bar the quadratic
sum of statistical and systematic errors. 
In most bins the inner errors are contained within the data point marker. 
Also shown are
the predictions from the DG model using GRV-LO real photon parton 
densities and  
$\omega=0.1~{\rm GeV}$ (solid line) and the SAS-1D (dashed line) 
and SAS-2D (dot-dashed line) parameterisations.}
\label{fig:epdfxg}
\end{center} 
\end{figure}
\newpage
\begin{figure}[ht]
\begin{center}
\epsfig{file=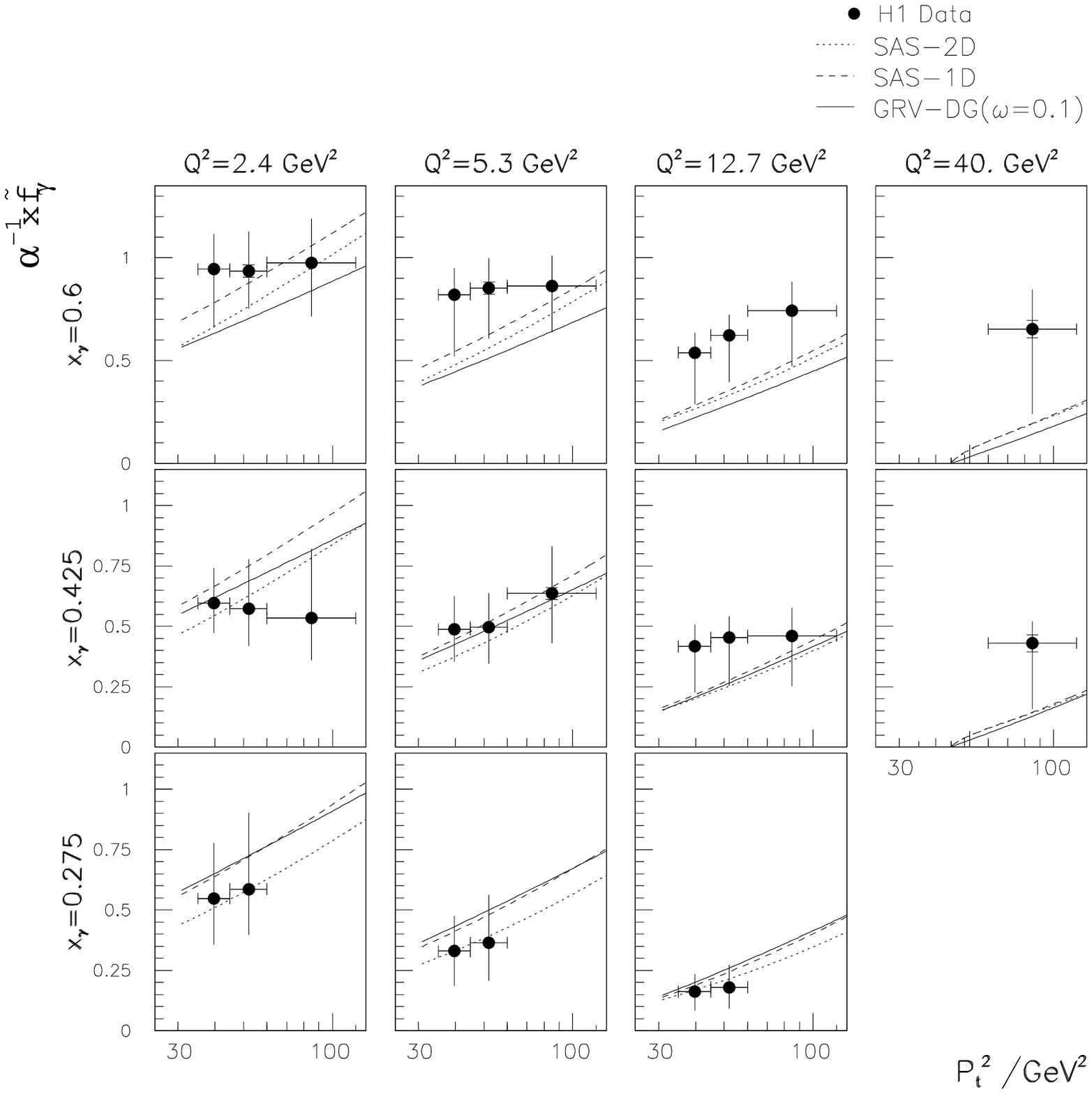,width=\textwidth}
\caption{The leading order effective parton density of the photon 
$\xgamma \tilde{f}_{\gamma} = 
\sum_{\rm n_f} \left(f_{{\rm q}/\gamma}
               +f_{\overline{\rm q}/\gamma}\right) +
\frac{9}{4} \,  f_{{\rm g}/\gamma}
$
, divided by the fine structure constant $\alpha$,
as a function of the squared parton transverse momentum, $P_t^2$,  
for different values of $Q^2$ and $x_\gamma$. 
The data are displayed as points, with the inner error bar depicting
the statistical error, and the total error bar the quadratic
sum of statistical and systematic errors. In most bins the inner errors are
contained within the data point marker. 
 Also shown are
the predictions from the DG model using GRV-LO real photon parton 
densities and  
$\omega=0.1~{\rm GeV}$ (solid line) and the SAS-1D (dashed line) 
and SAS-2D (dot-dashed line) parameterisations.}
\label{fig:epdfpt2}
\end{center} 
\end{figure}
\newpage
\begin{figure}[ht]
\begin{center}
\epsfig{file=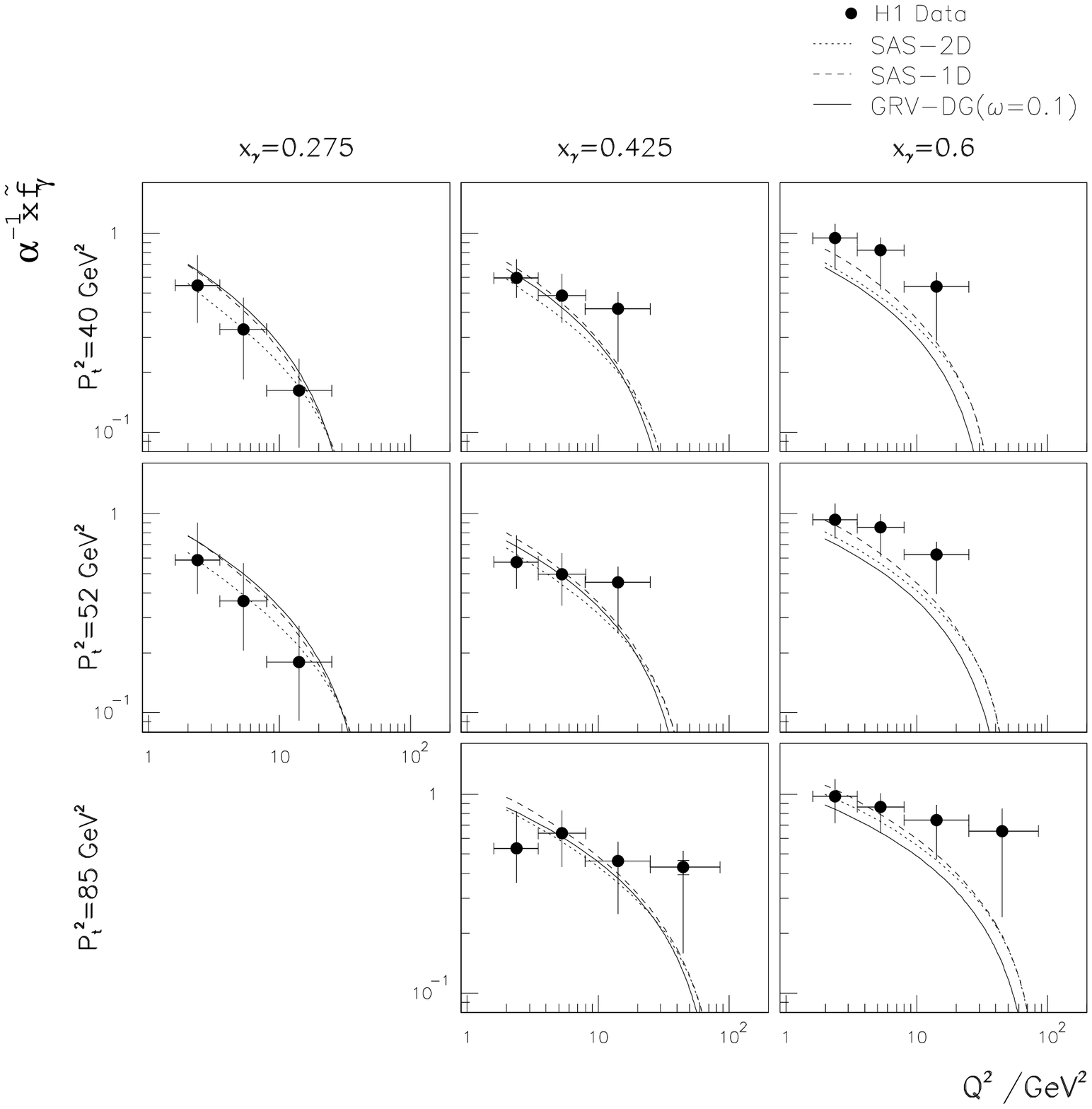,width=\textwidth}
\caption{The leading order effective parton density of the photon 
$\xgamma \tilde{f}_{\gamma} = 
\sum_{\rm n_f} \left(f_{{\rm q}/\gamma}
               +f_{\overline{\rm q}/\gamma}\right) +
\frac{9}{4} \,  f_{{\rm g}/\gamma}
$
, divided by the fine structure constant $\alpha$,
as a function of $Q^2$ for different values of $P_t^2$ and $x_\gamma$. 
The data are displayed as points, with the inner error bar depicting
the statistical error, and the total error bar the quadratic
sum of statistical and systematic errors. 
In most bins the inner errors are contained within the data point marker. 
Also shown are
the predictions from the DG model using GRV-LO real photon parton 
densities and  
$\omega=0.1~{\rm GeV}$ (solid line) and the SAS-1D (dashed line) 
and SAS-2D (dot-dashed line) parameterisations.}
\label{fig:epdfq2}
\end{center} 
\end{figure}
\newpage
\begin{figure}[ht]
\begin{center}
\epsfig{file=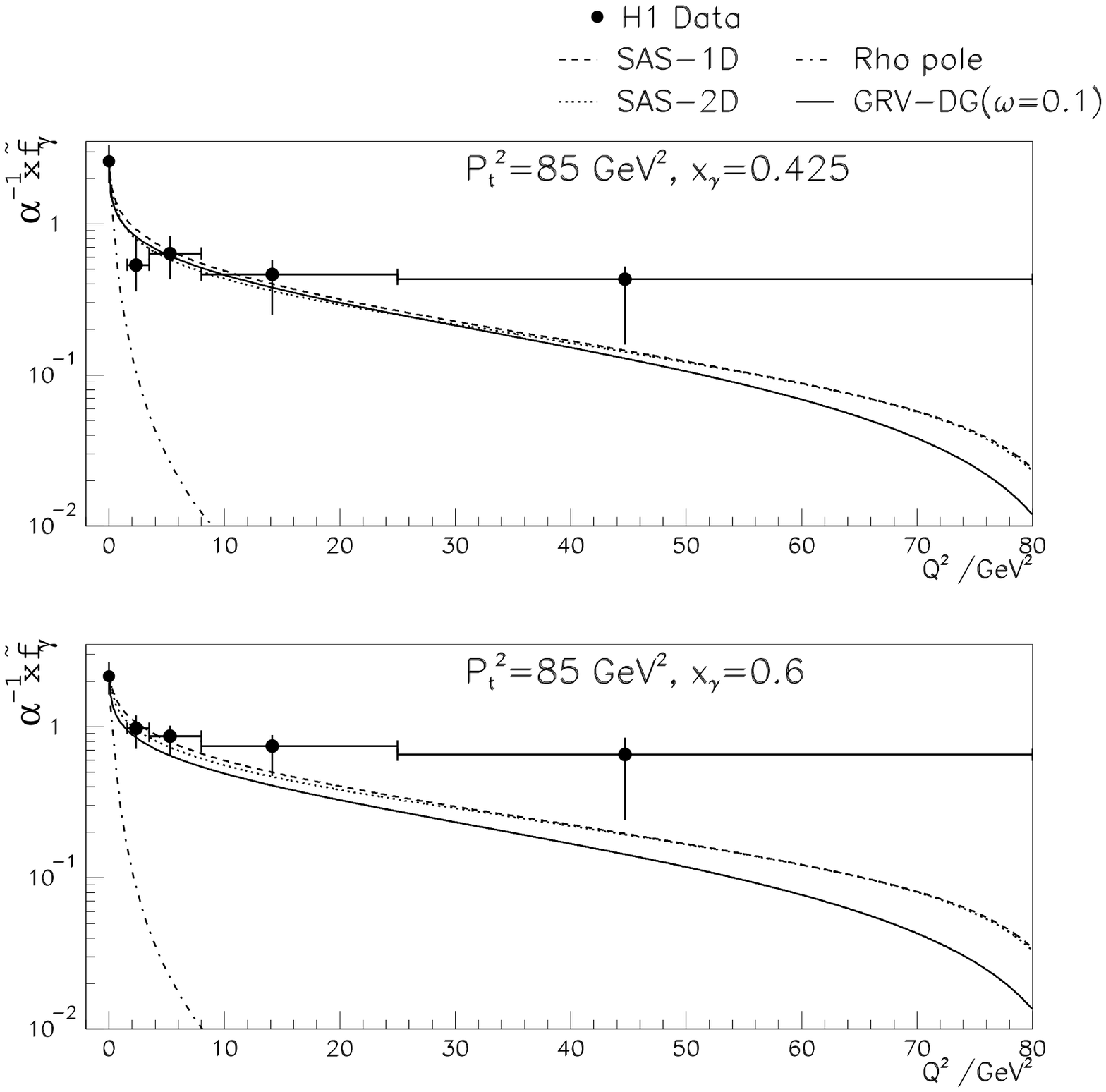,width=\textwidth}
\caption{The leading order effective parton density of the photon 
$\xgamma \tilde{f}_{\gamma} = 
\sum_{\rm n_f} \left(f_{{\rm q}/\gamma}
               +f_{\overline{\rm q}/\gamma}\right) +
\frac{9}{4} \,  f_{{\rm g}/\gamma}
$
, divided by the fine structure constant $\alpha$,
as a function of $Q^2$ for $P_t^2=85~{\rm GeV^2}$ and two
values of $\xgamma$.  
The data are displayed as points, with the inner error bar depicting
the statistical error, and the total error bar the quadratic
sum of statistical and systematic errors. 
The points at $Q^2=0$ are taken from
reference~\cite{Rick} and extrapolated to the right $P_t^2$ and $x_{\gamma}$
values by scaling with factors derived from 
GRV-LO parton densities for real photons.
Also shown are
the prediction from the DG model using GRV-LO photon parton 
densities and  
$\omega=0.1~{\rm GeV}$ (solid line) and the SAS-1D (dashed line) 
and SAS-2D (dotted line) parameterisations. 
The dot-dashed curve shows the photoproduction data scaled by a 
$\rho$-pole
factor (see text).}
\label{fig:epdfq22}
\end{center} 
\end{figure}

\begin{thebibliography}{99}
%
%
\bibitem{ggref1}
PLUTO Collab., Ch.\ Berger et al., Phys. Lett. {\bf B142} (1984) 111;\\
PLUTO Collab., Ch.\ Berger et al., Nucl. Phys. {\bf B281} (1987)  365;\\
PLUTO Collab., Ch.\ Berger et al., Phys. Lett. {\bf B107} (1981) 168;\\
JADE Collab., W.\ Bartel et al., Z. Phys. {\bf C24} (1984) 231; \\
TASSO Collab., M.\ Althoff et al.,  Z. Phys. {\bf C24} (1984) 231; \\
TPC/2$\gamma$ Collab., H.\ Aihara et al., Phys. Rev. Lett. {\bf 58} (1987) 97; \\
TPC/2$\gamma$ Collab., H.\ Aihara et al., Z. Phys. {\bf C34} (1987) 1; \\
%
AMY Collab., T.\ Sasaki et al., Phys. Lett. {\bf B252} (1990) 491;\\
OPAL Collab., R.\ Akers et al., Z. Phys. {\bf C61} (1994) 199;\\
AMY Collab., B.\ J.\ Kim et al., Phys. Lett. {\bf B325} (1994) 248;\\ 
TOPAZ Collab., H.\ Hayashii et al., Phys. Lett. {\bf B314} (1993) 149;\\
ALEPH Collab., D.\ Buskulic et al., Phys. Lett. {\bf B313} (1993) 509;\\
DELPHI Collab., P.\ Abreu et al., Phys. Lett. {\bf B342} (1995) 402.
%
\bibitem{herajets}
ZEUS Collab., M.\ Derrick et al., 
Phys. Lett. {\bf B322} (1994) 287; \\
H1 Collab., T.\ Ahmed et al.,
Nucl. Phys. {\bf B445} (1995) 195; \\
ZEUS Collab., M.\ Derrick et al., Phys. Lett. {\bf B384} (1996) 401; \\
ZEUS Collab., J.\ Breitweg et al.,  Eur. Phys. J. {\bf C1} (1998) 109.
%
\bibitem{Rick} 
H1 Collaboration, C. Adloff et al., Eur. Phys. J. {\bf C1} (1998) 97.
%
\bibitem{h1eflow} 
H1 Collab., S.\ Aid et al., Z. Phys. {\bf C70} (1996)   17.
%
%
%
\bibitem{vpth1}
T.\ Uematsu and T.\ Walsh, Phys. Lett. {\bf B101} (1981) 263; \\
T.\ Uematsu and T.\ Walsh, Nucl. Phys. {\bf B199} (1982) 93; \\
G.\ Rossi, U. California at San Diego preprint UCSD-10P10-227 (1983); \\
G.\ Rossi, Phys. Rev. {\bf D29} (1984) 852. 
\bibitem{ggmax}
PLUTO Collab., Ch.\ Berger et al., Phys. Lett. {\bf B142} (1984) 119.
%
\bibitem{vpth2}
M.\ Gl\"{u}ck, E.\ Reya and M.\ Stratmann,\
Phys. Rev. {\bf D54} (1996) 5515.
%
\bibitem{kkp}
M.\ Klasen, G.\ Kramer and B.\ P\"{o}tter, 
Eur. Phys. J. {\bf C1} (1998) 261.
%
\bibitem{vpth3}
D.\ de Florian, C.\ Garcia Canal and R.\ Sassot, Z. Phys. {\bf C75} (1997) 265.
%
\bibitem{vpth4}
J.\ Ch\'{y}la, J.\ Cvach, Proceedings of the Workshop 1995/96 on 
``Future Physics at HERA'', eds. G.\ Ingelman, A.\ de Roeck and R.\ Klanner,
DESY 1996, Vol. 1, 545.
%
\bibitem{vpth5}
M.\ Drees and R.\ Godbole, Phys. Rev. {\bf D50} (1994) 3124.
%
\bibitem{vpth6}
F.\ Borzumati and G.\ Schuler, Z. Phys. {\bf C58} (1993) 139.
%
%
\bibitem{loq2jet}
H1 Collaboration, C.\ Adloff et al., Phys. Lett. {\bf B415} (1997) 418.
%
\bibitem{marek}
 For detailed studies, refer to:\\
 M.Ta\v{s}evsk\'{y}, PhD Thesis, Prague, in preparation.
%
\bibitem{grind}
H1 Collaboration, {\it Di-Jet Event Rates in Deep-Inelastic Scattering at
  HERA}, DESY preprint 98-076, submitted to Eur. Phys. J. C.
%
\bibitem{inckt}
S.D.\ Ellis, D.E.\ Soper, Phys. Rev. {\bf D48} (1993) 3160;\\
S.\ Catani, Yu.L.\ Dokshitzer, M.H.\ Seymour and B.R.\ Webber, \
Nucl. Phys. {\bf B406} (1993) 187. 

%
%
%
\bibitem{h1det}
H1 Collaboration, I.\ Abt et al., Nucl. Instr. and Meth. 
{\bf A386} (1997) 310 and 348. 
%
\bibitem{spacal}
H1 SPACAL Group, R.\ D.\ Appuhn et al., Nucl. Instr. and Meth. 
{\bf A386} (1997) 397.
%
\bibitem{Lar}
H1 Calorimeter Group, B.\ Andrieu et al., Nucl. Instr. and Meth.
{\bf A336} (1993) 460.
%
\bibitem{tbeam}
H1 Calorimeter Group, B.\ Andrieu et al., Nucl. Instr. and Meth. 
{\bf A350} (1994) 57; \\
H1 Calorimeter Group, B.\ Andrieu et al., Nucl. Instr. and Meth. 
{\bf A336} (1993) 499.
%
%
%
\bibitem{herwig}
G.\ Marchesini et al., 
Comp. Phys. Comm. {\bf 67} (1992) 465.
%
%
\bibitem{rapgap}
H.\ Jung, Comp. Phys. Comm. {\bf 86} (1995) 147;\\
RAPGAP 2.06 manual, to be published.
%
\bibitem{herps}
G.\ Marchesini and B.\ Webber, Nucl. Phys. {\bf B238} (1984) 1; \\
G.\ Marchesini and B.\ Webber, Nucl. Phys. {\bf B310} (1988) 461.
%
\bibitem{jetset} 
T.\ Sj{\"o}strand, Comp. Phys. Comm. {\bf 82} (1994) 74; \\
T.\ Sj{\"o}strand, CERN-TH-7112-93 (Dec. 1993, revised Aug. 1994)  
%
%
\bibitem{GRVp}
M.\ Gl{\"u}ck, E.\ Reya and A.\ Vogt, Z. Phys. {\bf C67} (1995) 433.
%
\bibitem{z1.7}
ZEUS Collab., J.\ Breitweg et al.,  Eur. Phys. J. {\bf C1} (1998) 109.
%
\bibitem{GRVgam}
M.\ Gl{\"u}ck, E.\ Reya and A.\ Vogt, Phys. Rev. {\bf D45} (1992) 3986.
%
\bibitem{SAS}
T.\ Sj{\"o}strand and G.\ A.\ Schuler, Phys. Lett. {\bf B376} (1996) 193.
%
\bibitem{frixrid}
S.\ Frixione and G.\ Ridolfi,\
Nucl. Phys. {\bf B507} (1997) 315.
%
\bibitem{dagostini}
G.\ D'Agostini, Nucl. Instr. and Meth. {\bf A362} (1995) 487.
%
%
\bibitem{SES}
B.\ V.\ Combridge, C.\ J.\ Maxwell, Nucl. Phys. {\bf B239} (1984) 429. 
\bibitem{Budnev}
See for example, V.\ M.\ Budnev et al., Phys. Rep. {\bf C15} (1974) 181.
%
%
\bibitem{frixetal}
S.\ Frixione, M. \ Mangano, P.\ Nason and G.\ Ridolfi, 
Phys. Lett. {\bf B319} (1993) 339.

\end{thebibliography}
\end{document}